\documentclass[prc,twocolumn,superscriptaddress,showpacs,twoside,floatfix]
{revtex4}

\usepackage{graphicx}
\usepackage{amssymb}

\begin{document}

\title{Problems with interpretation of $^{10}$He ground state}

\author{L.\ V.\ Grigorenko}
\affiliation{Flerov Laboratory of Nuclear Reactions, JINR, RU-141980 Dubna,
Russia}
\affiliation{Gesellschaft f\"{u}r Schwerionenforschung
mbH, Planckstrasse 1,
D-64291, Darmstadt, Germany}
\affiliation{RRC ``The Kurchatov Institute'', Kurchatov sq.\ 1, 123182
Moscow, Russia}

\author{M.\ V.\ Zhukov}
\affiliation{Fundamental Physics, Chalmers University of Technology,
S-41296 G\"{o}teborg, Sweden}

\date{\today. {\tt File: /latex/10he/10he-17.tex }}

\begin{abstract}
The continuum of $^{10}$He nucleus is studied theoretically in a three-body 
$^{8}$He+$n$+$n$ model basing on the recent information concerning $^9$He 
spectrum [Golovkov, \textit{et al.}, Phys.\ Rev.\ C \textbf{76}, 021605(R) 
(2007)].  The $^{10}$He ground state (g.s.) candidate with structure 
$[p_{1/2}]^2$ for new g.s.\ energy of $^9$He is predicted to be at about 
$2.0-2.3$ MeV. The peak in the cross section associated with this state may be 
shifted to a lower energy (e.g.\ $\sim 1.2$ MeV) when $^{10}$He is populated in 
reactions with $^{11}$Li due to peculiar reaction mechanism. Formation of the 
low-energy ($E< 250$ keV) ``alternative'' ground state with structure 
$[s_{1/2}]^2$ is highly probable in $^{10}$He in the case of considerable 
attraction (e.g.\ $a<-5$ fm) in the $s$-wave $^9$He channel, which properties 
are still quite uncertain. This result either questions the existing 
experimental low-energy spectrum of $^{10}$He or place a limit on the scattering 
length in $^9$He channel, which contradicts existing data.
\end{abstract}

\pacs{21.60.Gx, 21.45.+v, 25.10.+s, 21.10.Dr}

\maketitle


\section{Introduction}


The first, at that moment theoretical, attempt to study $^{10}$He was undertaken 
in the end of 60-th \cite{baz69}. In this work a possibility of the 
nuclear-stable $^{10}$He existence was investigated in the microscopic 10-body 
hyperspherical harmonic (HH) model. However, until now the $^{10}$He nucleus 
remains relatively poorly studied system. Since it became clear that $^{10}$He 
is nuclear unstable \cite{ste88} and ground state properties of $^9$He were 
defined \cite{set87,boh88}, it became possible to predict theoretically the 
ground state of $^{10}$He as a narrow three-body $^8$He+$n$+$n$ resonance. It 
was found with $E \sim 0.7-0.9$, $\Gamma \sim 0.1-0.3$ MeV \cite{kor93}, for 
valence neutrons populating mainly $[p_{1/2}]^2$ configuration (the energy $E$ 
in the present work is always given relative to the three-body $^8$He+$n$+$n$ 
threshold). These predictions were soon confirmed experimentally: $E = 1.2(3)$, 
$\Gamma < 1.2$ MeV \cite{kor94}, $E = 1.07(7)$, $\Gamma = 0.3(2)$ MeV 
\cite{ost94,boh99}, and  $E = 1.7\pm 0.3 \pm 0.3$ MeV \cite{kob97}.

A new possible theoretical understanding of $^{10}$He was proposed after the 
existence of a virtual state in $^9$He was demonstrated by Chen \textit{et al}.\ 
in Ref.\ \cite{che01}. An \emph{upper} limit for scattering length $a<-10$ fm 
was established in this experimental work. For  such an attractive $s$-wave 
interaction in $^9$He Aoyama  predicted in Ref.\ \cite{aoy02} the existence of a 
narrow near-threshold $0^+$ state in $^{10}$He ($E = 0.05$, $\Gamma = 0.21$ MeV) 
with the $[s_{1/2}]^2$ structure  in addition to the $[p_{1/2}]^2$ state 
(calculated in this work to be at about 1.7 MeV). Concerning evident discrepancy 
with the experimental data the author of Ref.\ \cite{aoy02} suggested that the 
ground state (g.s.) of $^{10}$He had not been observed so far and the state at 
$\sim 1.3$ MeV is actually the first excited state. However, no possible 
explanation was proposed in Ref.\ \cite{aoy02} for which reason the 
$[s_{1/2}]^2$ g.s.\ was missed in experiments.

In recent experiment by Golovkov \textit{et al}.\ \cite{gol07} at Dubna 
radioactive beam facility ACCULINNA the low-lying spectrum of $^9$He was 
revised, providing a higher position of the $p_{1/2}$ state than in the previous 
studies. A broad $p_{1/2}$ state was observed at about 2 MeV  instead of the 
(presumably) $p_{1/2}$-$p_{3/2}$ doublet of narrow states at 1.27 and 2.4 MeV as 
in Refs.\ \cite{set87,boh88,boh99}. The experiment \cite{gol07} also claims a 
unique spin-parity identification below 5 MeV. The presence of the $s_{1/2}$ 
contribution is evident in the data \cite{gol07}, but the exact nature of this 
contribution is still unclear, whether it is a virtual state with considerably 
large negative scattering length or just a smooth nonresonant background. A 
relaxed \emph{lower} limit for scattering length $a>-20$ fm was established in 
this work. These new data should have a strong impact on the calculated 
properties of $^{10}$He, which inspired us to ``revisit'' the issue.

We study the question in theoretical models, which are schematic but have a 
clear relevance to real possible reaction mechanisms of the $^{10}$He continuum 
population. In a contrast with approach of Ref.\ \cite{aoy02}, which provided 
only energies and widths of the states, we are interested in the observable 
consequences of the $J^{\pi}=0^+$ states with structures $[s_{1/2}]^2$ and 
$[p_{1/2}]^2$ ``coexistence'' in the $^{10}$He spectrum. We demonstrate that 
this problem has a key importance for understanding of observable properties of 
$^{10}$He. We arrive to a conclusion that the simultaneous consistent 
understanding of the low-lying spectra of $^9$He and $^{10}$He is still a 
challenge both from theoretical and experimental sides.

The unit system $\hbar=c=1$ is used in this work.


\section{Theoretical model}


To choose the interactions in this work we generally follow the prescription of
the three-cluster $^8$He+$n$+$n$ calculations of Ref.\ \cite{kor93} with 
appropriate modifications of potentials. From the set of the core-$n$ potentials 
tested in \cite{kor93} we selected one (denoted there as ``I2''). Other choices 
do not change qualitatively the result and quantitatively are quite close. The 
potential is parameterized by Gaussian formfactor
\[
V^l_{c,ls}(r) =  V^l_{c,ls} \exp[-r^2/r^2_0]
\]
with $r_0=3.4$ fm. The depths of the $d$-wave potential $V^2_c = -33$ MeV and 
the $(ls)$ component in $p$-wave $V^1_{ls} = 10$ MeV, are the same as in 
original paper. The inverse $(ls)$ forces were used in Ref.\ \cite{kor93} in 
$p$-wave to account for occupied $p_{3/2}$ subshell in the $^8$He core. The 
interaction in the $s$-wave $^8$He-$n$ channel was pure repulsive in Ref.\ 
\cite{kor93} to account for an occupied $s_{1/2}$ shell in the $^8$He core. 
Central potential parameters in $s$- and $p$-waves $V^0_c$ and $V^1_c$ are
being varied to clarify different aspects of the system dynamics. To manage the 
occupied $s_{1/2}$ state in $^8$He in this work an additional repulsive core is 
introduced in the $s$-wave with parameters $r_0(core)=2.35$ fm and 
$V^0_c(core) = 75$ MeV. 

With the above potential the $d$-wave state in $^9$He is found at 4.8 MeV
which is consistent with the the experimental data \cite{set87,boh99,gol07} 
giving values in the range $4.2-4.9$ MeV for the $d_{5/2}$ state. With $V^1_c = 
-10$ MeV (the value from the Ref.\ \cite{kor93}) the $p_{1/2}$ state is obtained 
at 0.74 MeV. This value is different from value 1.15 MeV quoted in Ref.\ 
\cite{kor93}, where this is the energy at which the phase shift pass $\pi/2$. In 
this work we have to deal also with broad states, where the phase shift does not 
reach  $\pi/2$. Thus, we define the resonance position for two-body subsystem by 
``observable value'' (peak in the elastic cross section) and define the width as 
the full width on half maximum (FWHM) for this peak. 

The realistic soft-core potential \cite{gog70} is used in the $n$-$n$ subsystem
also following Ref.\ \cite{kor93}.

\begin{figure}
\centerline{\includegraphics[width=0.47\textwidth]{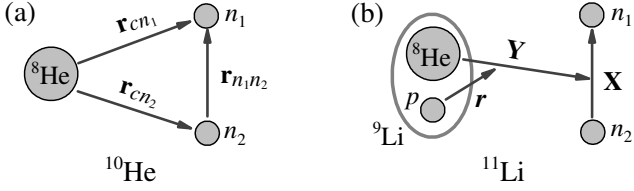}}
\caption{Coordinate sets used in this paper. Panel (b) illustrates a
proton removal from $^{11}$Li as a method to populate $^{10}$He.}
\label{fig:coor}
\end{figure}

To study qualitatively a possible influence of the reaction
mechanism we follow the approach of paper \cite{gri03b} to exotic
$^5$H system. We introduce a compact source function
$\Phi(\rho,\Omega_{\rho})$ in the right hand side of the three-body
Schr\"{o}dinger equation and solve the inhomogeneous system of
equations
\begin{eqnarray}
\left(\hat{H}-E \right) \Psi_E^{(+)}(\rho,\Omega_{\rho}) =
\Phi(\rho,\Omega_{\rho}) \;,
\label{eq:source} \\
\hat{H} = \hat{T} + \hat{V}_{cn}(\mathbf{r}_{cn_1}) +
\hat{V}_{cn}(\mathbf{r}_{cn_2})+ \hat{V}_{nn}(\mathbf{r}_{n_1n_2})\;, \nonumber
\\
\rho^2 = \frac{8}{10}\left( r^2_{cn_1}+r^2_{cn_2} \right) +
\frac{1}{10}r^2_{n_1n_2}  = \frac{1}{2}X^2+\frac{8}{5}Y^2 \;,
\end{eqnarray}
for pure outgoing wave boundary conditions, utilizing the
hyperspherical harmonic (HH) method. The used coordinates are shown
in Fig.\ \ref{fig:coor}. The hyperradial components
$\chi^{(+)}_{K\gamma}(\rho)$ of the WF
\[
\Psi_E^{(+)}(\rho,\Omega_{\rho}) = \rho^{-5/2} \sum ^{K_{\max}}_{K \gamma}  \,
\chi^{(+)}_{K \gamma}( \varkappa \rho) \,
\mathcal{J}_{K \gamma}^{JM}(\Omega _{\rho}) \;  ,
\]
are matched to Riccati-Bessel functions of half-integer index
$\mathcal{H}^{(+)}_{K+3/2}$. Functions $\mathcal{H}^{(+)}$ have the
asymptotic behavior  $\exp[i \varkappa \rho]$, where
$\varkappa=\sqrt{2ME}$ ($M$ is a nucleon mass), describing the
partial outgoing waves for hyperspherical equations. The value
$K_{\max}$ truncates the hyperspherical expansion.  The hypermoment
$\varkappa$ is expressed via the energies of the subsystems $E_x$,
$E_y$ or via Jacobi momenta $k_x$, $k_y$ conjugated to Jacobi
coordinates $X$, $Y$:
\begin{eqnarray}
 \mathbf{k}_x & = &  \frac {1} {2} \left(  \mathbf{k}_{n_1} - \mathbf{k}_{n_2}
\right) \; ,\quad
\mathbf{k}_y  =  \frac{4}{5} \left( \mathbf{k}_{n_1} + \mathbf{k}_{n_2} \right)
- \frac{1}{5} \mathbf{k}_{c} \;,
\nonumber \\
\varkappa^2   & = &2ME = 2M(E_x+E_y) = 2 k_x^2 + \frac{5}{8}\,k_y^2 \;.
\label{eq:momenta}
\end{eqnarray}
The Jacobi variables are given in ``T'' Jacobi system. A more
detailed picture of Jacobi coordinates for coordinate and momentum
spaces in ``T'' and ``Y'' Jacobi systems can be found in Fig.\
\ref{fig:corel}.

The set of coupled equations for functions $\chi^{(+)}$ has the form
\begin{eqnarray}
\left[  \frac{d^{2}}{d\rho^{2}}-\frac{\mathcal{L}(\mathcal{L}+1)}{\rho^{2}}+
2M\left\{  E-V_{K \gamma,K \gamma}(\rho)\right\}  \right]  \chi^{(+)}_{K \gamma}
(\rho) \qquad \nonumber \\
= 2M \sum_{K' \gamma '} V_{K \gamma,K^{ \prime
}\gamma^{\prime}}(\rho)\chi^{(+)}_{K^{\prime}
\gamma^{ \prime }}(\rho) + 2M \,\Phi_{K \gamma}(\rho) \, ,
\label{shredl} \\
V_{K\gamma,K^{\prime}\gamma^{\prime}}(\rho)=\int \!\! d \Omega_{\rho} \,
\mathcal{J}_{K' \gamma'}^{JM*}(\Omega_{\rho}) \sum_{i<j}
V_{ij}(\mathbf{r}_{ij})\,\mathcal{J}_{K \gamma}^{JM}(\Omega_{\rho})\, ,
\label{hhpot} \\
\Phi_{K \gamma}(\rho) = \rho^{5/2} \int d \Omega_{\rho} \,
\mathcal{J}_{K \gamma}^{JM*}(\Omega_{\rho}) \, \Phi(\rho,\Omega_{\rho})  \, ,
\nonumber
\end{eqnarray}
where ${\cal L}=K+3/2$ and $V_{K\gamma,K^{\prime}\gamma^{\prime}}(\rho)$ are
matrix elements of the sum of the pairwise  potentials referred to in this work
as three-body potentials.

More detailed account of the method can be found e.g.\ in Ref.\
\cite{gri03b}. It is shown there that the method is consistent with
``sudden removal'' approximation for high energy fragmentation
reactions. The development of the technically similar approach in
the framework of the DWBA theory, applied to the inelastic processes in the 
transfer reactions, can be found in Ref.\ \cite{asc69}.

We used two different sources, consistent with different reaction
conditions. One is a ``narrow'' source with a Gaussian formfactor
\begin{equation}
\Phi(\rho,\Omega_{\rho}) =  \exp[-\rho^2/\rho_0^2] \;
\sum_{K=0,2 \; S=0} \mathcal{J}_{K \gamma}^{JM}(\Omega _{\rho}) \;,
\label{eq:narrow}
\end{equation}
where $\rho_0=4.1$ fm provides the source rms radius $\langle \rho
\rangle = 5$ fm. This is a typical radius for the ``reaction
volume'' for ordinary nuclei. The source populates only the lowest
hyperspherical components of the WF ($K=0,2$). This qualitatively
corresponds to the population of the $[s]^2$ and $[p]^2$ shell model
configurations in the $^{10}$He nucleus, which are expected to be
the most important for the low-energy part of the spectrum. The
condition $S=0$ is qualitatively consistent with mechanism of
transfer reactions, where the $^{10}$He states are populated by
transferring a two-neutron pair (with total spin equal zero) to the
$^8$He core.
In such reactions the $\Delta S=1$ transfer is strongly suppressed
and the $\Delta S=0$ transfer is a very reliable assumption.

The other choice of the source is more reaction specific. When
$^{10}$He is produced from $^{11}$Li in a process which can be
approximated as a sudden proton removal from $^{9}$Li core, the
source term $\Phi(\rho,\Omega_{\rho})$ should contain the Fourier
transform of the overlap integral between the $^8$He WF
$\Psi_{^{8}\mbox{\scriptsize He}}$, the spin-isospin function of the
removed proton $\chi_p$ and the $^{11}$Li wave function over the
radius-vector $\mathbf{r}$ between the removed proton and the
center-of-mass of $^{10}$He [see Fig.\ \ref{fig:coor} (b)]:
\begin{equation}
\Phi(\rho,\Omega_{\rho}) =
\int d \mathbf{r} \, e^{i \mathbf{qr}} \langle \Psi_{^{8}\mbox{\scriptsize He}}
\chi_p| \Psi_{^{11}\mbox{\scriptsize Li}} \rangle \, .
\label{eq:overlap}
\end{equation}
In general, this quantity is a complicated function of the recoil
momentum vector $\mathbf{q}$, transferred to the $^{10}$He system in
the proton removal process. However, if the reaction energy is large
and the internal energy of $^{10}$He is small, one can neglect this
dependence (see Ref.\ \cite{gri03b} for details).
It can be shown that in this case partial hyperspherical components
of the source function are well approximated by the corresponding
components of the $^{11}$Li WF. Thus, this type of calculations is
further referred as ``$^{11}$Li source''. The $^{11}$Li WF was
taken from an analytical parametrization developed in Ref.\
\cite{shu06} taking into account broad range of experimental
information on this nucleus. The dominant $[s_{1/2}]^2$ and
$[p_{1/2}]^2$ configurations are populated by the $^{11}$Li source
with almost equal probabilities. The rms radius of such a source
function $\langle \rho \rangle = 9.5$ fm is enormous compared to
typical nuclear sizes.

In the approach with the source function of Eq.\ (\ref{eq:source}) the cross 
section for population of the $^{10}$He continuum is proportional to the 
outgoing flux of the three particles on a hypersphere of some large radius $\rho 
= \rho_{\max}$:
\begin{equation}
\frac{d \sigma}{dE} \sim \frac{1}{M} \mathop{\rm Im} \!
\int \! d \Omega _{\rho } \left. \Psi_E ^{(+)\dagger}
\rho^{5/2}\frac{d}{d\rho} \, \rho^{5/2} \,
\Psi_E^{(+)}\right| _{\rho =\rho_{\max} } \, .
\label{eq:cross}
\end{equation}
Differentials of this flux on the hypersphere provide angular and
energy distributions among the decay products at given decay energy
$E$ (see Ref.\ \cite{gri03b} for details of correlation
calculations).


\section{Calculations}


\subsection{Basis size convergence}


\begin{figure}
\includegraphics[width=0.42\textwidth]{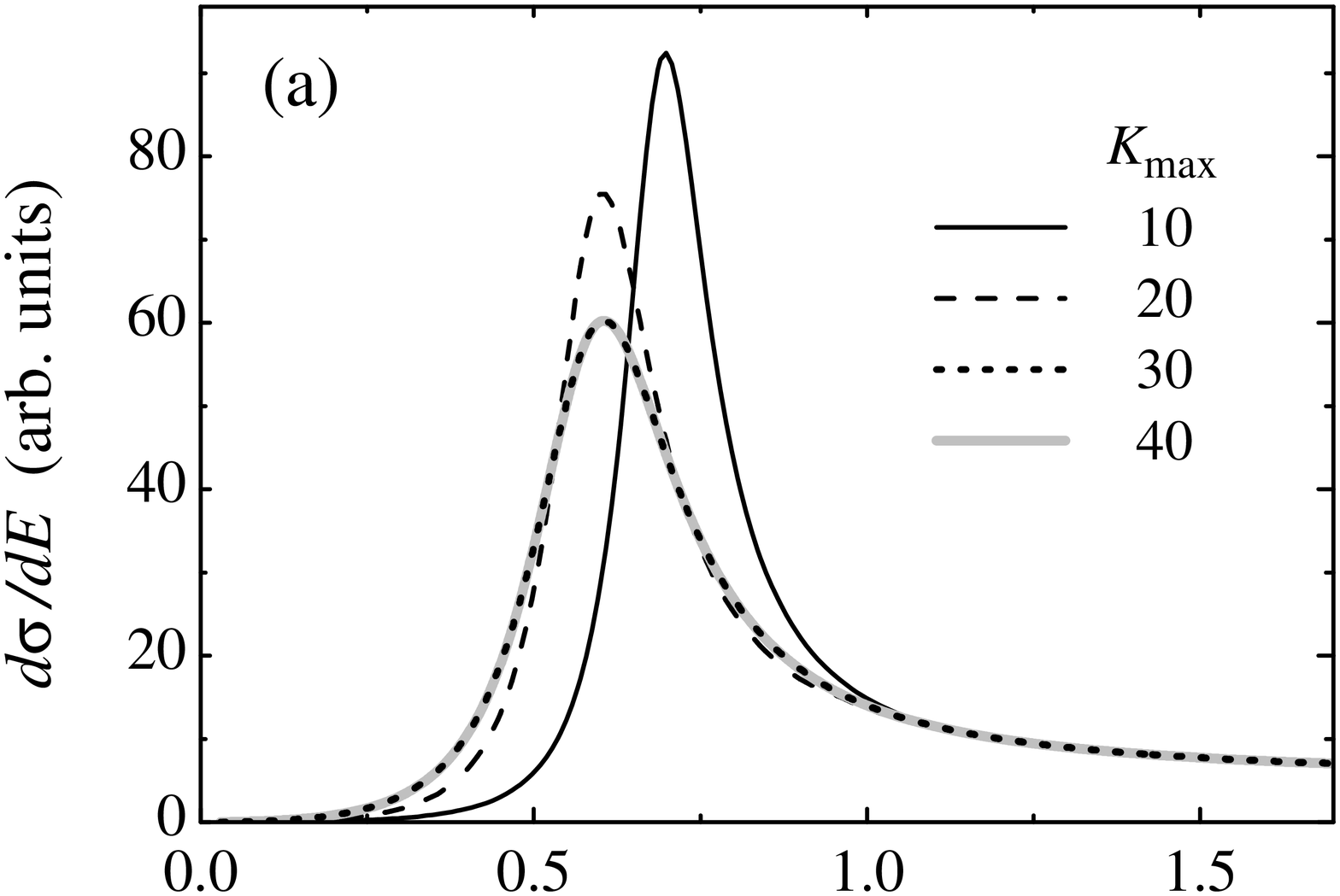} \\
\includegraphics[width=0.425\textwidth]{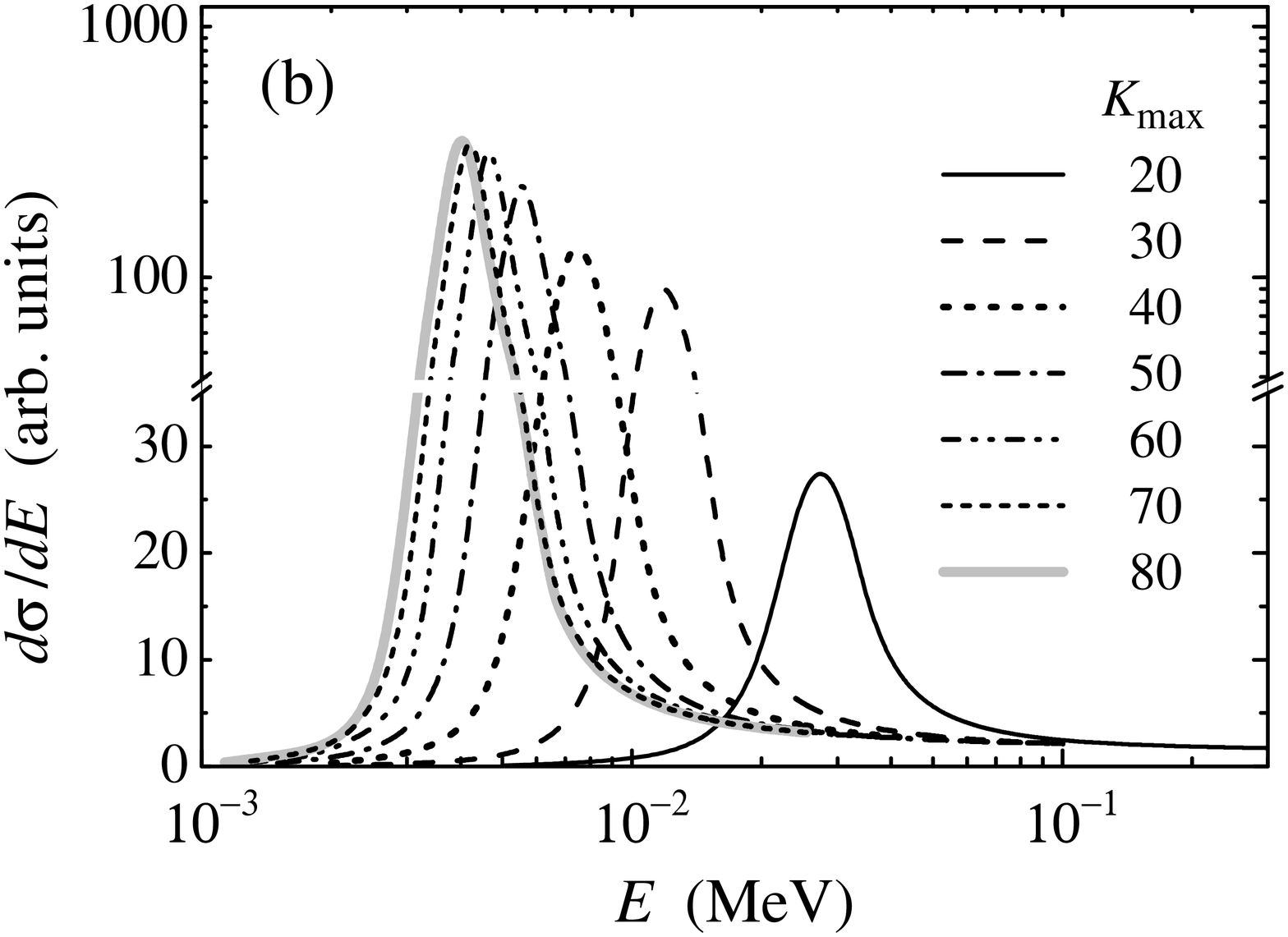}
\caption{Convergence of calculations as a function of $K_{\max}$
(the value truncating the hyperspherical basis). Calculations with
narrow source. (a) Resonance peak with $[p_{1/2}]^2$ structure. The
$^8$He-$n$ potential parameters are: $V^0_c=0$, $V^1_c=-10$ MeV. (b)
Resonance peak with $[s_{1/2}]^2$ structure. Parameters are:
$V^0_c=-26.93$ MeV (this corresponds to $a=-15$ fm in $^9$He),
$V^1_c=-4.5$ MeV.}
\label{fig:hh-conv}
\end{figure}

The HH calculations in our method can be performed with
$K_{\max}=24-26$. Such basis sizes could be not sufficient for
obtaining a good energy convergence of calculations in some
complicated cases. The basis size can be further increased
effectively using the adiabatic procedure based on the so called
Feshbach reduction (FR) \cite{gri07}. Feshbach reduction eliminates
from the total WF $\Psi=\Psi_{p}+\Psi_{q}$ an arbitrary subspace $q$
using the Green's function of this subspace:
\[
H_{p}=T_{p}+V_{p}-V_{pq}G_{q}V_{pq} \;.
\]
In a certain adiabatic approximation we can assume that the radial
part of kinetic energy is small under the centrifugal barrier in the
channels with high centrifugal barriers and can be approximated by a
constant. In this approximation the FR procedure is reduced to the
construction of effective three-body interactions
$V^{\text{eff}}_{K\gamma,K^{\prime}\gamma^{\prime}}$ by matrix
operations
\begin{eqnarray}
G_{K\gamma,K^{\prime}\gamma^{\prime}}^{-1} & = & (H-E)_{K\gamma,K^{\prime
}\gamma^{\prime}}=V_{K\gamma,K^{\prime}\gamma^{\prime}} \nonumber \\
 & + & \left[  E_{f}-E+\frac{(K+3/2)(K+5/2)}{2M\rho^{2}}\right]
\delta_{K\gamma,K^{\prime} \gamma^{\prime}}\,, \nonumber \\
V^{\text{eff}}_{K\gamma, K^{\prime}\gamma^{\prime}} & = &
V_{K\gamma,K'\gamma'}-\sum V_{K\gamma,\bar{K}\bar{\gamma}}
G_{\bar{K}\bar{\gamma}, \bar{K}^{\prime} \bar{\gamma}^{\prime}}
V_{\bar{K}^{\prime}\bar{\gamma}^{\prime}, K^{\prime
}\gamma^{\prime}}\;.\nonumber
\end{eqnarray}
Summation over indexes with bar is made for eliminated channels. We take the
``Feshbach energy'' $E_{f}$ in our calculations as $E_{f}\equiv E$.

\begin{figure}
\includegraphics[width=0.44\textwidth]{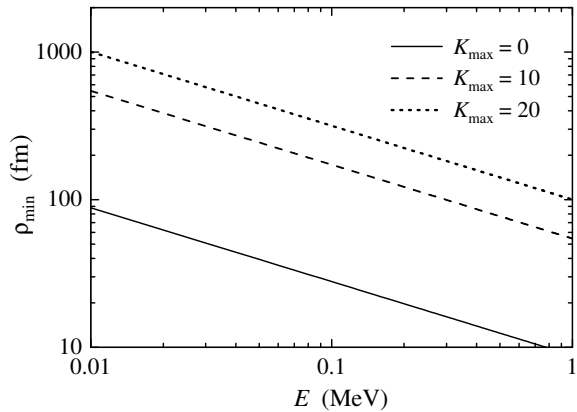}
\caption{Hyperradius of the classical turning point $\rho_{\min}$  for
hyperradial centrifugal barriers in the channels with different $K$ values.}
\label{fig:rhomin}
\end{figure}

Reliability of the FR procedure can be checked in two ways. We can compare 
dynamic calculations for some large $K_{\max}$ with the ``reduced'' calculations 
$K_{\max} \rightarrow K_{FR}$ (with much smaller dynamic basis size $K_{FR}$) 
and in principle they should coincide. Calculations show that for $^{10}$He 
starting from $K_{\max}=26$ we get practically the same result down to 
$K_{FR}=10$. The other way to make a check is the following. We can also start 
FR with some quite large fixed $K_{\max}$ (e.g.\ from $K_{\max}=100$ in this 
work), make the reduction to different $K_{FR} \leq 26$ and perform dynamic 
calculations with each of them. Again the results were found to coincide 
precisely for $K_{FR} \geq 10$. Thus it was found reliable to perform most of 
the calculations (except those for correlations) with dynamic basis size $K_{FR} 
= 12$ varying effective basis size $K_{\max}$ when necessary.

The cross section convergence with the increase of the effective hyperspherical 
basis size is demonstrated in Fig.\ \ref{fig:hh-conv}. For resonance peak with 
$[p_{1/2}]^2$ structure the convergence is reliably achieved by $K_{\max}=30$. 
However, in the case of a state with $[s_{1/2}]^2$ structure more efforts are 
required to achieve the convergence (very close to the threshold even a minor 
variation of the energy becomes noticeable). We intentionally demonstrate in 
Fig.\ \ref{fig:hh-conv} (b) the case, which is numerically more complicated than 
the others considered in the paper. When the $s$-wave potential in the 
$^8$He-$n$ subsystem is taken to provide the scattering length $a=-15$ fm, the 
resonance peak in $^{10}$He is obtained at $E=4$ keV with $\Gamma=0.7$ keV. The 
basis size $K_{\max}=80$ is required in such a case to obtain the convergence.

Another aspect of the basis size choice is connected with the radial extent of 
the calculations $\rho_{\max}$. The formulation of the cross section 
calculations in the form (\ref{eq:cross}) implies that the WF residues at 
$\rho_{\max}$ in the classically allowed region. Taking into account the 
character of the hyperspherical centrifugal barrier (\ref{shredl}) this requires 
a very large radial extent for large basis sizes. Fig.\ \ref{fig:rhomin} 
provides the estimates of the minimally required values of $\rho_{\max}$ to 
satisfy this condition for different $K$ values. So, we used $\rho_{\max} \sim 
300-500$ fm for calculations of the $[p_{1/2}]^2$ states and $\rho_{\max} \sim 
1000-2000$ fm for the extreme low-energy calculations of the $[s_{1/2}]^2$ 
states.

\begin{figure}
\includegraphics[width=0.42\textwidth]{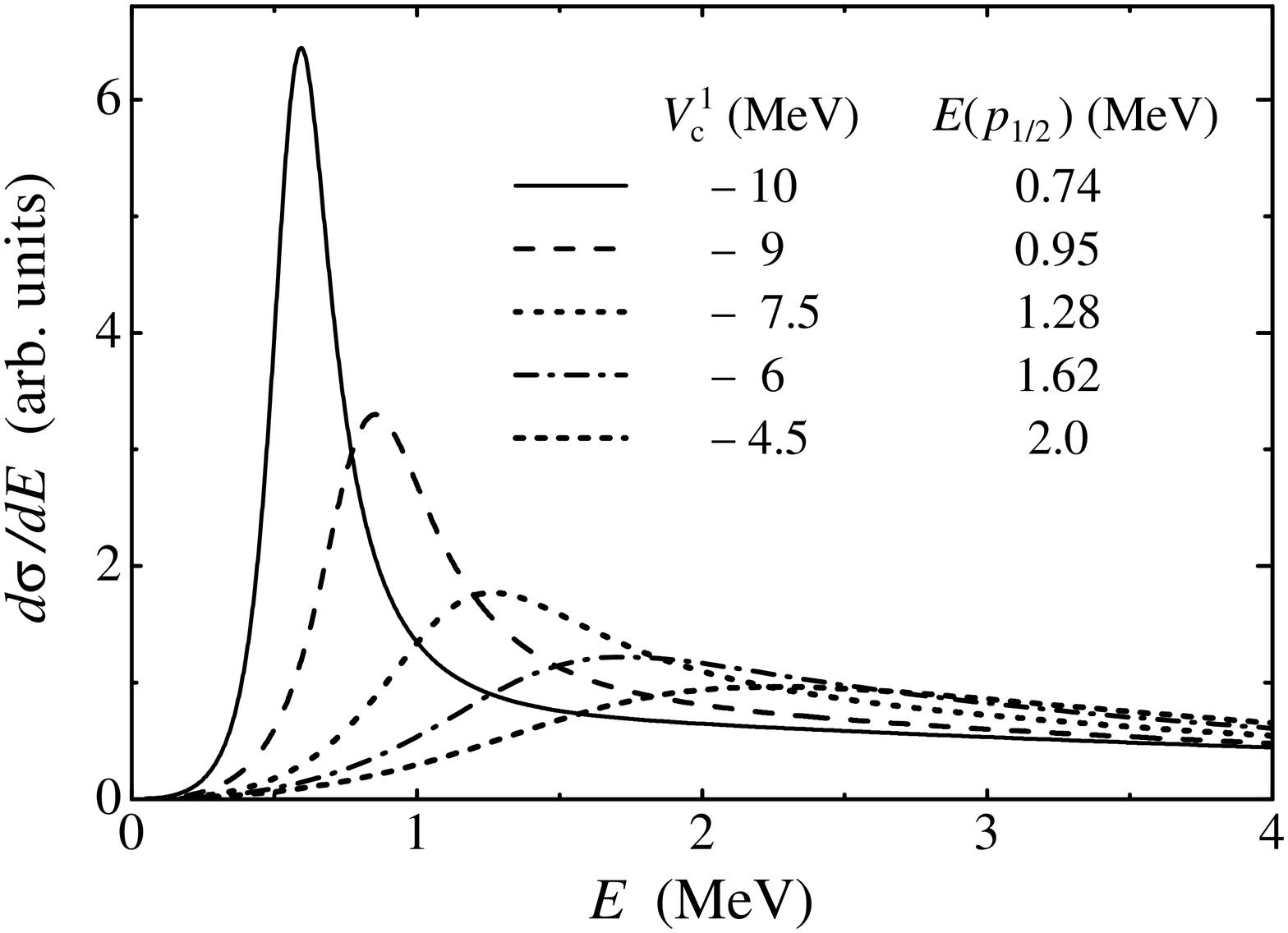}
\caption{Behavior of the $^{10}$He spectrum with decrease of the $p$-wave 
potential depth $V^1_c$. The corresponding $p_{1/2}$ state energies $E(p_{1/2})$ 
in $^9$He relative to the $^8$He-$n$ threshold are shown in the legend. 
Calculations with narrow source.}
\label{fig:p-dec}
\end{figure}


\subsection{Sensitivity to the $p$-wave in $^9$He}


The ground state resonance properties were predicted as $E \sim 0.7-0.9$ MeV, 
$\Gamma \sim 0.1-0.3$ MeV in Ref.\ \cite{kor93}. Within the approach used in 
this work we first of all reproduce the results of previous studies. The 
calculation with model parameters consistent with these of Ref.\ \cite{kor93} is 
shown in Fig.\ \ref{fig:p-dec}, by solid curve. The peak position is somewhat 
lower than in Ref.\ \cite{kor93} ($E=0.6$ MeV, $\Gamma = 0.27$ MeV) which is 
connected to the larger basis size (see Fig.\ \ref{fig:hh-conv}(a)). Note that 
$K_{\max}=8$ was used in Ref.\ \cite{kor93}.

\begin{figure*}
\includegraphics[width=0.342\textwidth]{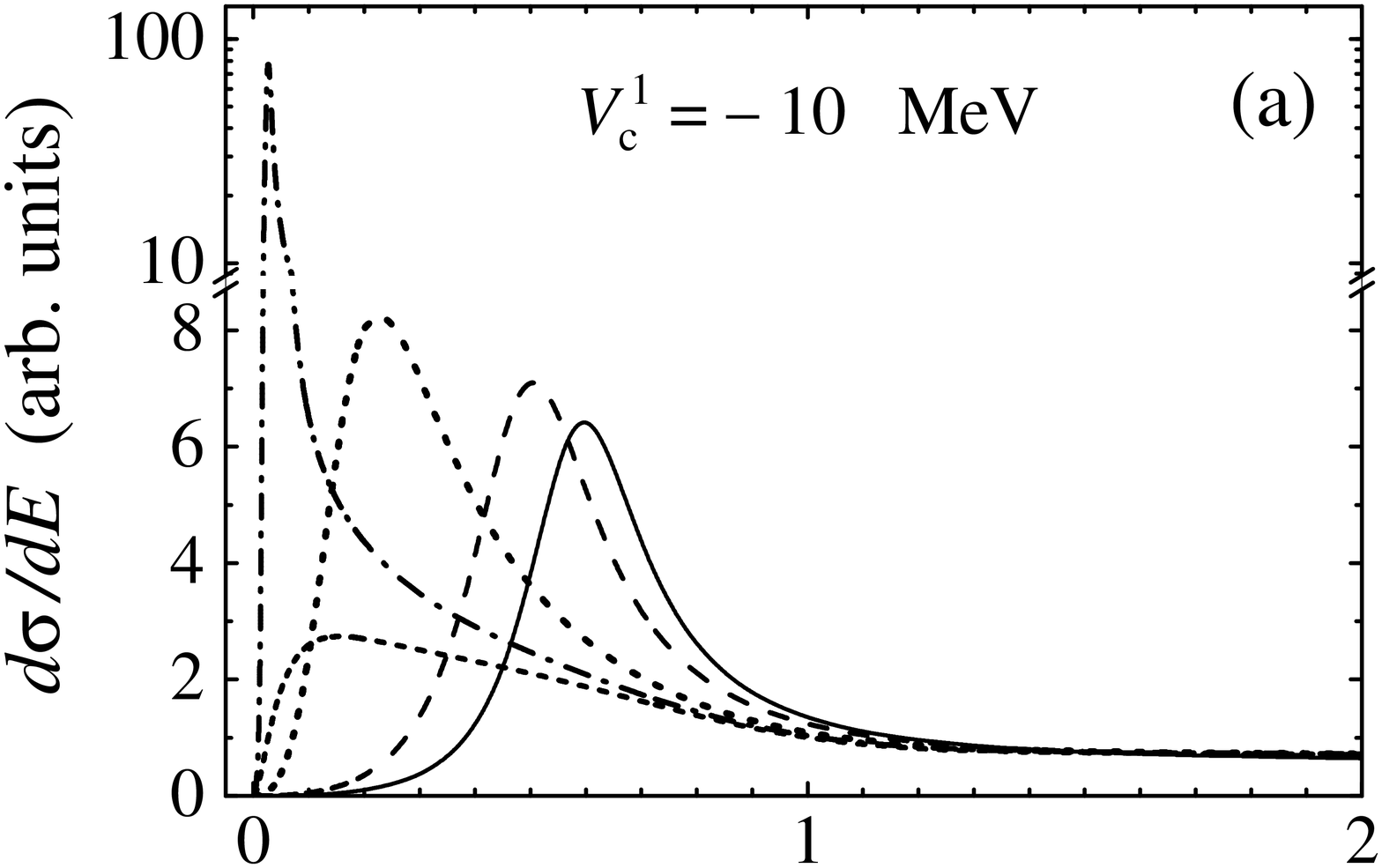}
\includegraphics[width=0.328\textwidth]{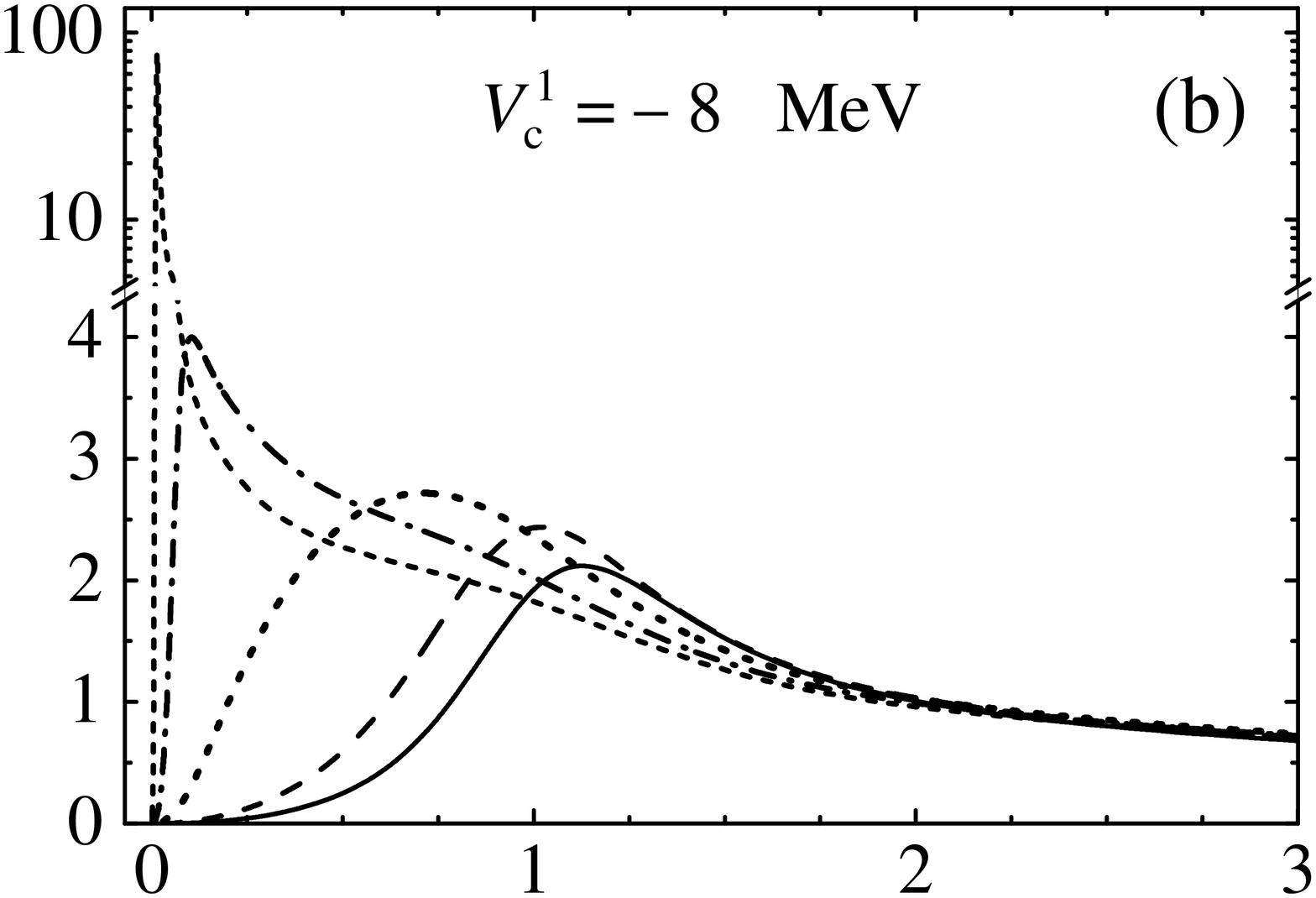}
\includegraphics[width=0.312\textwidth]{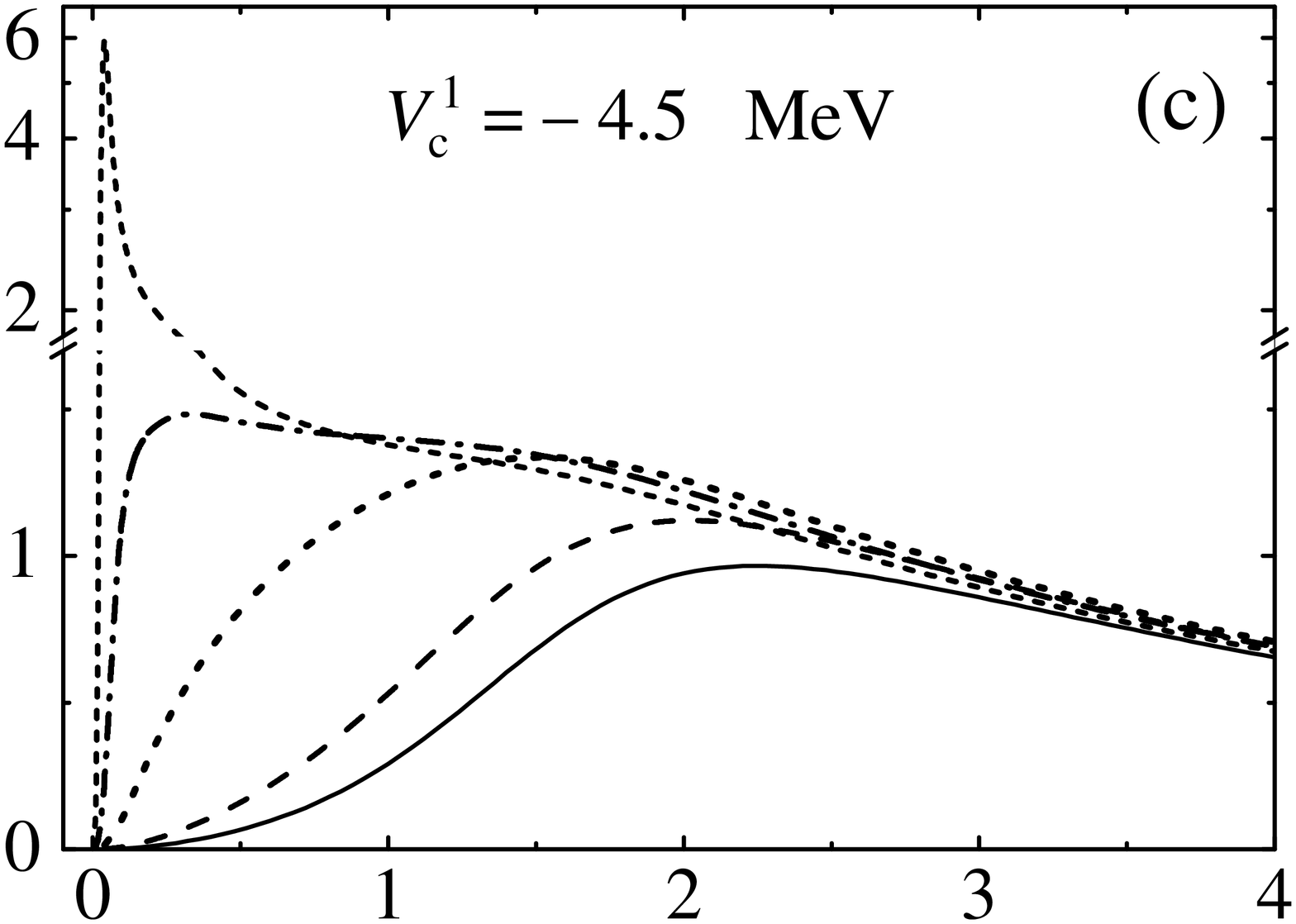} \\
\includegraphics[width=0.342\textwidth]{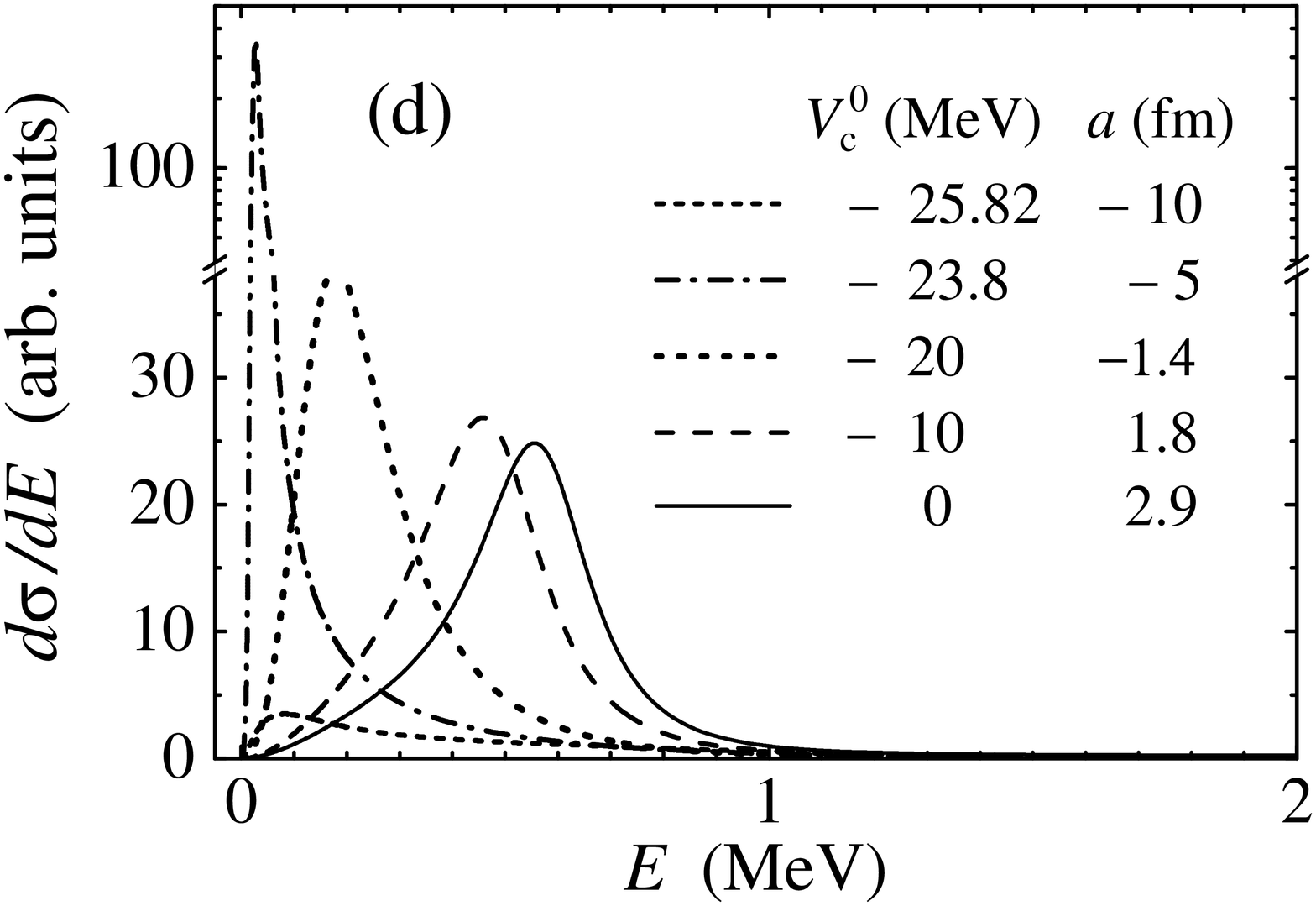}
\includegraphics[width=0.323\textwidth]{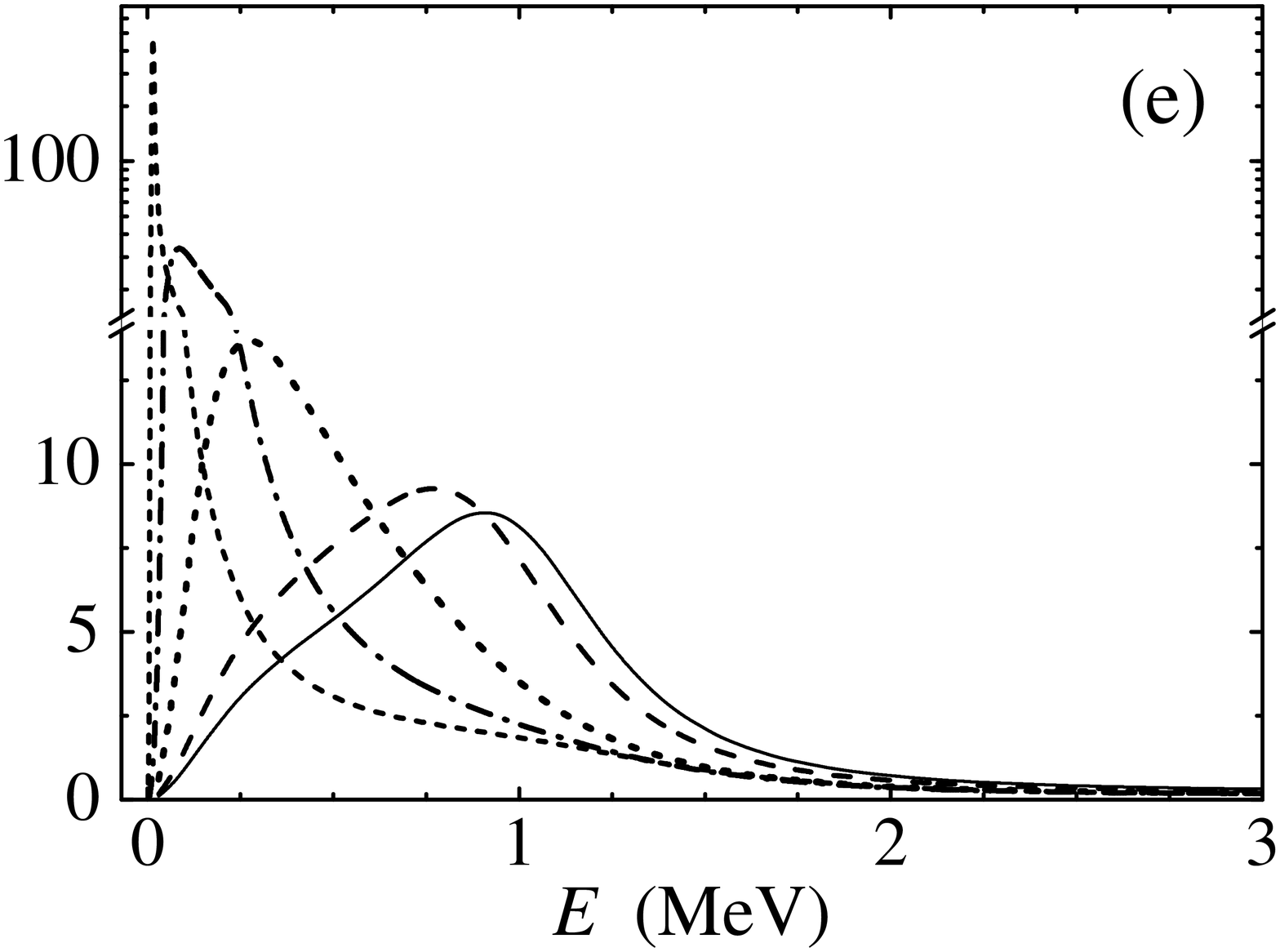}
\includegraphics[width=0.323\textwidth]{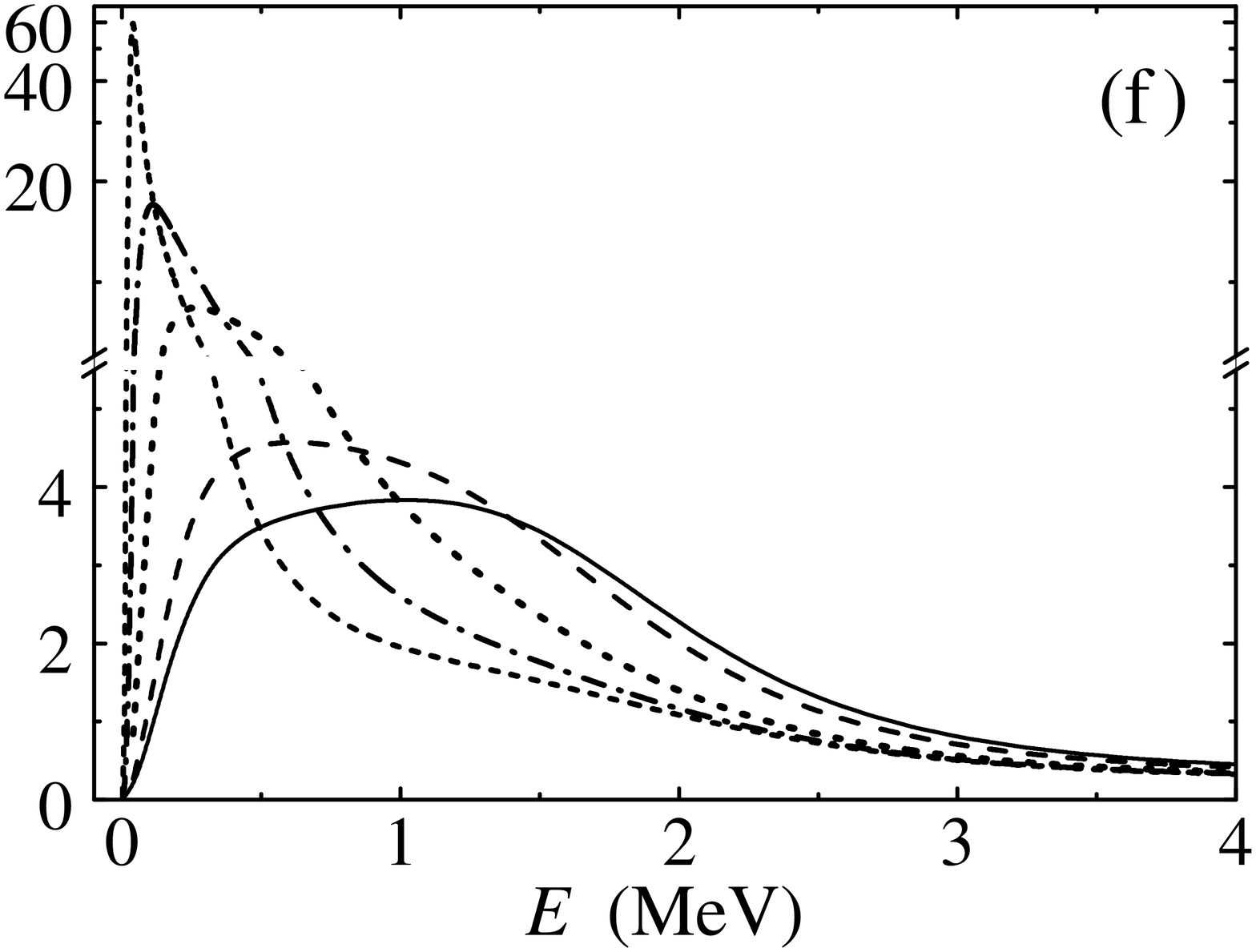} \\
\caption{Behavior of the $^{10}$He spectrum with increase of the
$s$-wave interaction [legend, the same for all panels is shown in
panel (d)] . The first row shows the narrow source case, the second
row is for broad ($^{11}$Li) source. In the first column
calculations for $p$-wave potential $V^1_c=-10$ MeV ($p_{1/2}$ state
at 0.74 MeV, as in calculations of Ref.\ \cite{kor93}); in the
second column for $V^1_c=-8$ MeV ($p_{1/2}$ state at 1.16 MeV, close
to 1.27 MeV, as in experiment \cite{boh99}); in the third column for
$V^1_c=-4.5$ MeV ($p_{1/2}$ state at 2 MeV, as in experiment
\cite{gol07}). Smooth behavior of the short-dashed curves
($V^0_c=-25.82$ MeV, $a=-10$ fm) in panels (a) and (d) is connected
to the fact that the bound $0^+$ state of $^{10}$He is formed with
binding energy $\approx 60$ keV. Note, the change of the scales on the
vertical axes to logarithmic.}
\label{fig:s-inc}
\end{figure*}

The evolution of the cross section with decrease of the $p$-wave
interaction from the value adopted in Ref.\ \cite{kor93}
($V^1_c=-10$ MeV, which provided the energy of the $p_{1/2}$ state
$E(p_{1/2})=0.74$ MeV) to a value providing the $^9$He g.s.\ to be
at about 2 MeV ($V^1_c=-4.5$ MeV), is shown in Fig.\ \ref{fig:p-dec}
for the narrow source function. The new peak position for the
$^{10}$He population cross section is at $E=2.3$ MeV. The impact of
this change is drastic: the narrow $^{10}$He g.s.\ peak is
practically ``dissolved'' as the system becomes less bound: e.g.\
the width of the peak can not be any more well defined as FWHM.


\subsection{Sensitivity to the reaction mechanism}


\begin{figure}
\includegraphics[width=0.38\textwidth]{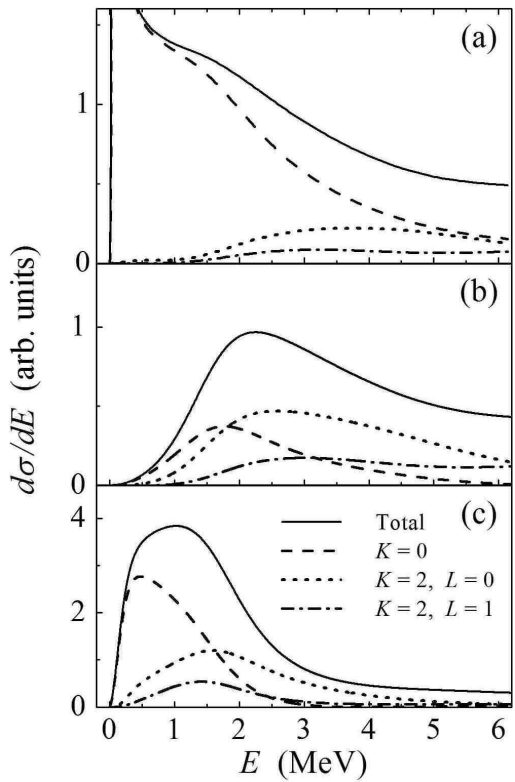}
\caption{Partial decomposition of the cross section; dashed, dotted and
dash-dotted curves provide contributions of the main WF components. Calculations
with $p$-wave resonance in $^9$He at 2 MeV ($V^1_c=-4.5$ MeV). Different panels
correspond to: (a) $V^0_c=-25.82$ ($a=-10$ fm), narrow source; (b) $V^0_c=0$
narrow source, and (c) $V^0_c=0$, $^{11}$Li source.}
\label{fig:bn-com}
\end{figure}

The evolution of cases with different $p$-wave interactions with increase of the
$s$-wave interaction is shown in Fig.\ \ref{fig:s-inc} for the narrow and broad
source functions, which should simulate different reaction conditions. We first
discuss sensitivity of the cross section to the reaction mechanism.

The narrow ground state in $^{10}$He is not significantly sensitive to the 
reaction mechanism. This can be seen comparing Figs.\ \ref{fig:s-inc} (a) and 
(d): difference of curves of the same style in the upper and lower panels is 
quantitative, not qualitative. This is an expected result as the narrow states 
have a sufficiently large lifetime to ``forget'' how they were populated and 
thus loose the sensitivity to the population mechanism.

When the state is above 1 MeV, the width becomes comparable to 1 MeV and the 
dependence on the source function is considerable [Figs.\ \ref{fig:s-inc} (b) 
and (e)]. In the case of even higher $^{10}$He g.s.\ the calculations with 
narrow and broad sources have very little in common [Figs.\ \ref{fig:s-inc} (c) 
and (f)]. According to the recent result \cite{gol07} the cases (c) and (f) 
should be regarded as the most realistic. Thus peculiarities of the reaction 
mechanism could be a problem for interpretation of the $^{10}$He spectra.


\subsection{Sensitivity to the $s$-wave in $^9$He}


It can be seen from Fig.\ \ref{fig:s-inc} that for relatively weak $s$-wave 
attraction the g.s.\ peak is shifted to lower energies with minimal distortion. 
However, as the $s$-wave attraction becomes stronger the threshold peculiarity 
is formed in the spectrum. With the further increase of the $s$-wave interaction 
this peculiarity is transformed into very sharp low-energy ($E< 300$ keV) peak. 
The WF at this peak has a practically pure $[s_{1/2}]^2$ structure and we 
characterize it as a ``three-body virtual state''.

Using the term ``three-body virtual state'' we have two things in mind: this is 
an $s$-wave state build upon the virtual states in all the subsystems, and this 
state has distinct properties compared to ordinary resonant three-body states 
(relevant discussion of ``Efimov-like three-body virtual excitations'' can be 
found in Ref.\ \cite{dan07}). 

The ordinary two-body virtual states are typically 
characterized in two ways: (i) as a negative energy pole on the second Riemann 
sheet or (ii) as a threshold peculiarity \footnote{For example in the inelastic 
process the virtual state is seen as near threshold peak with non-Lorencian 
shape.} preceding the formation of the bound state in the case of absence of the 
potential barrier.  The pole behavior in the three-body systems has been studied 
in a number of works with the emphasis on the possible similarities with 
two-body virtual state poles behavior \cite{glo78,tan99,del00,aoy02,fre07}. In 
paper \cite{fre07} the possibility of such behavior in the three-body $s$-wave 
system was shown for interactions with certain extreme properties. 
Observable consequences of such a pole behavior in the three-body systems remain 
unclear. Our way to think about three-body virtual state is more relevant to the 
second characteristic of the two-body virtual state, which is connected to its 
observables. 

For relatively strong $s$-wave interaction in the $^8$He-$n$ subsystem (namely 
such that the scattering length $a<-5$ fm), we unavoidably (means independently 
on the structure and reaction mechanism details) get a sharp peak in the cross 
section with energy less than 0.3 MeV and with dominating $[s_{1/2}]^2$ 
configuration. Stable formation of the low-energy peak at certain strength of 
attractive $s$-wave interaction in $^9$He is an important dynamical feature of 
the $^{10}$He system which makes us optimistic about predictive abilities of 
theoretical models in this situation. The extreme low-energy peaks could hardly 
be consistent with the experimental data \cite{kor94}, the discussion of the 
issue is provided below in Section \ref{sec:exp-consist}.

It can be noticed that in the case of the very narrow three-body virtual state 
formation, some structure can be seen as a ``shoulder'' on the right slope of 
the  $[s_{1/2}]^2$ peak. It is possible to understand that this structure 
corresponds to the state with the $[p_{1/2}]^2$ structure which becomes 
sufficiently well split from the $[s_{1/2}]^2$ state and even preserves the 
position typical for $V_c^0=0$ case. The analysis of the partial decomposition 
of the cross section provided in Fig.\ \ref{fig:bn-com} indicates that this is 
generally true. However, the $[p_{1/2}]^2$ contribution to WF is considerably 
broadened and reduced in absolute value compared to the case when there was no 
$s$-wave attraction. For understanding of Fig.\ \ref{fig:bn-com} it is useful to 
note that at the ``shell model language'' the $K=0$ configuration is a pure 
$[s_{1/2}]^2$, while the $K=2$ components (for $p$-shell nuclei) are mainly 
decomposed as
\begin{eqnarray}
\left| K=2,L=0 \right\rangle  & = & \sqrt{1/3} \; [p_{1/2}]^2  + \sqrt{2/3} \;
[p_{3/2}]^2 \;,
\nonumber \\
\left| K=2,L=1 \right\rangle  & = & \sqrt{2/3} \; [p_{1/2}]^2 - \sqrt{1/3} \;
[p_{3/2}]^2 \;.
\nonumber
\label{eq:wf-decomp}
\end{eqnarray}
The weight of the $[p_{1/2}]^2$ configuration  relative to the total weight of
$[p_{1/2}]^2$ and $[p_{3/2}]^2$ configurations varies from 80 to 90 percent in
different calculations of the $p$-wave state.


\subsection{Properties of the three-body virtual state}


Important feature which differs a three-body virtual state (the one with 
dominant $[s_{1/2}]^2$ structure) from the ordinary two-body virtual states is 
evident from the structure of equations (\ref{shredl}). This feature has been 
exploratory discussed in the past (e.g.\ Ref.\ \cite{tan99}), but it seems that 
in $^{10}$He this kind of physics could become really accessible for 
observation. The state with $[s_{1/2}]^2$ structure should be characterized by 
domination of component with lowest possible value of the generalized angular 
momentum $K=0$. However, the centrifugal barrier 
$\mathcal{L}(\mathcal{L}+1)/2M\rho^2$ is not zero even in the channel with 
$K=0$, as it depends on ``effective angular momentum'' $\mathcal{L}=K+3/2$. This 
means that the low-energy three-body virtual state may exist in the form of a 
real resonance peak, not a threshold peculiarity as two-body virtual state. It 
is also easy to demonstrate that the low-energy behavior of the inelastic cross 
section for population of the three-body continuum is
\[
d \sigma / dE \propto E^2 \;,
\]
in contrast with the two-body inelastic cross section which has a square root 
peculiarity in the case of the virtual state
\[
d \sigma / dE \propto \sqrt{E}\;.
\]
Such a behavior should in principle distinctly separate the three-body virtual 
state peak from zero energy. Such a separation was demonstrated in Ref.\ 
\cite{tan99} for a toy model of the $[s^2]$ state for the ``Borromean system'' 
\footnote{What is called the ``Borromean'' property of resonances, due to 
relation to artificially created Borromean states in ACCC method \cite{tan99}, 
is better characterized as energy condition of a ``true three-body decay'' or a 
``democratic decay''. The detailed discussion of three-body decay modes can be 
found in Ref.\ \cite{gri01}.}. Namely, it was shown in the analytical 
continuation of coupling constant (ACCC) method that the pole trajectories in 
the case of the $[s^2]$ three-body state are analogous to the trajectories in 
the system with barriers, while for the two-body virtual states they are 
qualitatively different.

\begin{figure}
\includegraphics[width=0.48\textwidth]{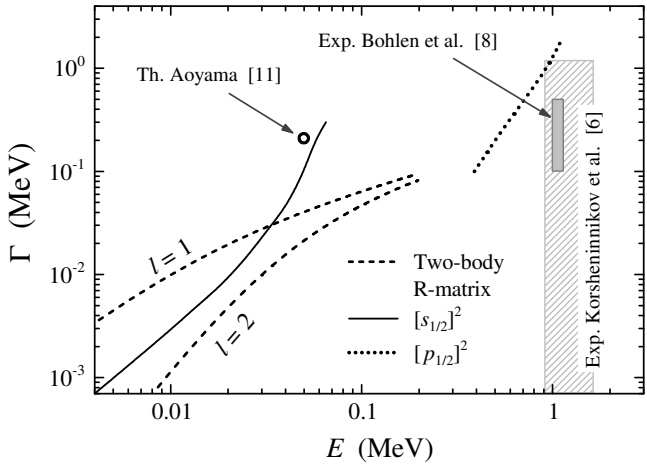}
\caption{Width as a function of resonance energy for $[s_{1/2}]^2$
and $[p_{1/2}]^2$ states. Standard two-body R-matrix estimates are
shown for $l=1$ and $l=2$ (channel radius 40 fm) by dashed curves.
Possible experimental ranges according to experiments
\cite{kor94,boh99} are shown by hatched and grey rectangles
correspondingly. The result of theoretical prediction \cite{aoy02}
is shown by a small circle. The curve for $[p_{1/2}]^2$ state is
calculated with the broad source. Calculations for $[s_{1/2}]^2$
state with broad and narrow sources practically coincide within the
shown energy range.}
\label{fig:gam-ot-e}
\end{figure}

\begin{figure*}
\includegraphics[width=0.93\textwidth]{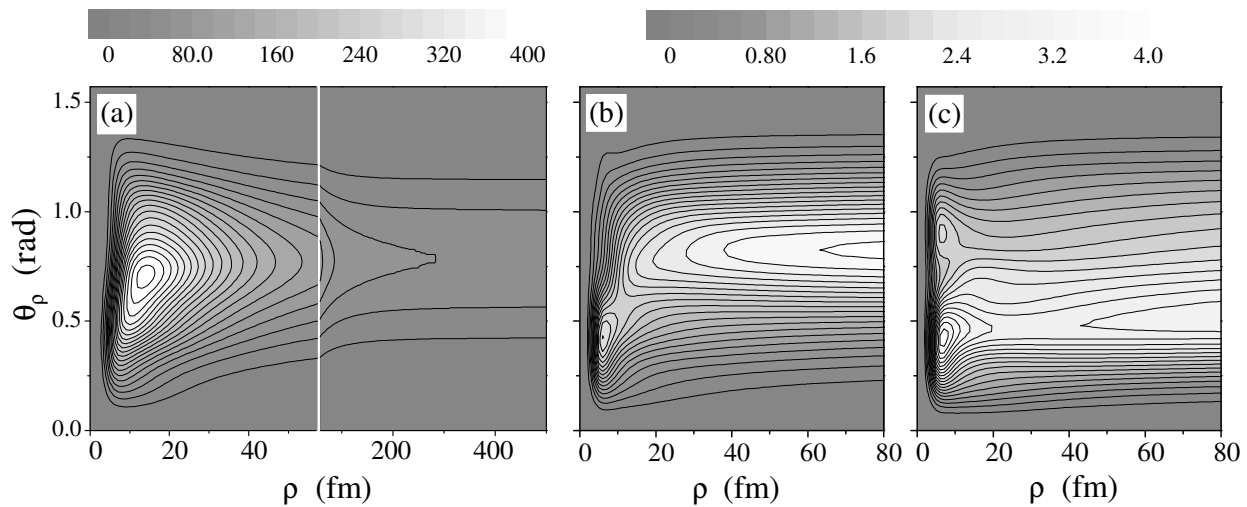}
\caption{Correlation density $|\Psi^{(+)}|^2$ for the $^{10}$He WF on the
$\{\rho,\theta_{\rho}\}$ plane in the ``T'' Jacobi system. Panel (a) shows case
of strong three-body virtual state $V^0_c=-26.93$ ($a=-15$ fm), $V^1_c=-4.5$
MeV for energy taken on peak position ($E=4$ keV). See also Fig.\
\ref{fig:hh-conv} (b). Panel (b) gives the same as (a) but for $E=2.3$
MeV (the latter energy is the expected position for peak in the case of state
with $[p_{1/2}]^2$ structure). Panel (c) shows the state with $[p_{1/2}]^2$
structure $V^0_c=0$, $V^1_c=-4.5$ MeV. Calculations are made with narrow source;
energy is taken on peak position ($E=2.3$ MeV).}
\label{fig:wfden}
\end{figure*}

The mentioned features, however, do not mean that a three-body virtual state is 
an ordinary resonance state; there is an important difference. It is known that 
the resonance behavior is connected to time delays in the propagation of 
particles (and corresponding time-dependent theory can be formulated in these 
terms). Ordinarily, the time delay is connected to the confinement of particles 
inside the potential barrier and their WF is localized inside the potential 
well, displaying the ``quasibound'' nature of such resonances. In the case of 
virtual state the time delay is not connected with barrier and tight spacial 
localization of particles close to each other. It is connected with \emph{slow} 
motion of particles in the volume of sphere with \emph{large} radius (comparable 
to scattering length). In the three-body case the hyperspherical centrifugal 
barrier $\mathcal{L}(\mathcal{L}+1)/2M\rho^2$ has an effective collective 
nature; it is clear that individual nucleons in $[s_{1/2}]^2$ configuration do 
not ``see'' any barriers. The time delay is connected therefore to the 
simultaneous presence of two valence nucleons in the volume around the core, 
associated with scattering lengths, which means a peripheral nature of such a 
state.

The peripheral character of the state presumes different character of the 
dependence of the resonance width on energy than the behavior which could be 
expected for ``typical'' barrier penetration. We can take, for example, the 
calculated properties of the $^{10}$He g.s.\ in the case $a=-15$ fm ($E=4$ keV, 
$\Gamma=0.7$ keV, see Fig.\  \ref{fig:hh-conv}) and deduce the ``channel 
radius'' $\rho_{\text{ch}}$ using for penetration expression analogous to the 
single-chanel R-matrix formula. It can be found in Ref.\ \cite{gol04a}:
\begin{equation}
\Gamma = \frac{1}{M \rho^2_{\text{ch}}} \; \frac{2}{\pi} \;
\frac{1}{J^2_{K+2}(\varkappa \rho_{\text{ch}})+N^2_{K+2}(\varkappa
\rho_{\text{ch}})} \;.
\label{gamm-3}
\end{equation}
Then the value $\rho_{\text{ch}} \approx 40$ fm is obtained. So, the radial 
range, which can be interpreted as an ``internal region'' in the case of the 
virtual three-body state, is huge.

The dependence of the width (defined as FWHM) on resonance energy is shown in 
Fig.\ \ref{fig:gam-ot-e} for $[s_{1/2}]^2$ and $[p_{1/2}]^2$ resonance peaks. 
The variation of energy in each case is obtained by respective variation of 
parameters of the $s$- and $p$-wave interactions. The $[p_{1/2}]^2$ curve is 
obtained in calculations with a broad source. The curves for $[s_{1/2}]^2$ 
calculations practically coincide for broad and narrow source calculations; this 
independence is an expected result for such a narrow structure. It can be seen 
in Fig.\ \ref{fig:gam-ot-e} that the curve for $[s_{1/2}]^2$ state stays mainly 
in between standard two-body R-matrix estimates with $l=1$, $l=2$ (channel 
radius 40 fm) as an ``effective angular momentum'' for $K=0$ is 
$\mathcal{L}=3/2$. The obtained dependence is in a good agreement with the 
$^{10}$He ground state prediction by Aoyama \cite{aoy02} which gave $E=0.05$ MeV 
and $\Gamma=0.21$ MeV (small circle in Fig.\ \ref{fig:gam-ot-e}). We think that 
this agreement is an important fact demonstrating stability of the theoretical 
results on this issue because very different theoretical models and different 
$p$-wave interactions were employed in the studies of Ref.\ \cite{aoy02}.

The point about a peripheral character of the $[s_{1/2}]^2$ state is also 
confirmed by analysis of the correlation density. The correlation densities 
$|\Psi^{(+)}|^2$ for the $^{10}$He WFs on the $\{\rho,\theta_{\rho}\}$ plane are 
shown in Fig.\ \ref{fig:wfden}. The $\theta_{\rho}$ hyperspherical variable 
describes the distribution between $X$ and $Y$ subsystems. It is a component of 
the 5-dimensional hyperangle $\Omega_{\rho}=\{\theta_{\rho},\Omega_x, \Omega_y 
\}$. For $^{10}$He in the ``T'' Jacobi system
\[
X=\sqrt{2} \, \rho \sin \theta_{\rho} \quad ; \quad
Y=\sqrt{5/8} \, \rho \cos \theta_{\rho} \;.
\]
Some properties of the three-body virtual state are well illustrated by this
plot.

\begin{enumerate}

\item The $[s_{1/2}]^2$ and $[p_{1/2}]^2$ configurations demonstrate very 
different correlations in the internal region and on asymptotic. While in the 
$[s_{1/2}]^2$ case the distributions are expectedly quite featureless, in the 
$[p_{1/2}]^2$ case we observe in the internal region the double-humped 
structures --- ``dineutron'' and ``cigar'' --- in the variable $\theta_{\rho}$, 
which are connected to so called Pauli focusing [Fig.\ \ref{fig:wfden} (b), (c); 
in the case (b) only ``dineutron'' peak is seen]. These structures  are well 
known from the studies of the other $p$-shell nuclei \cite{zhu93}. In the case 
when there is no attractive $s$-wave interaction in the $^8$He-$n$ channel 
[Fig.\ \ref{fig:wfden} (c)] this double-humped correlation ``survives'' up to 
the asymptotic region in somewhat modified form and thus could possibly be 
observed in experiment (see also the discussion in the Section \ref{sec:corel}).

\item Peripheral character of the $[s_{1/2}]^2$ WF. It can be seen from a
comparison of Figs.\ \ref{fig:wfden} (b) and (c) that the double-humped 
structure connected to $[p_{1/2}]^2$ configurations is sharply concentrated in 
the internal region ($\rho \sim 7$ fm) and rapidly decreases beyond $10-15$ fm. 
In contrast, the $[s_{1/2}]^2$ WF is peaked at $\rho \sim 15$ fm and extends 
smoothly to around 50 fm [Fig.\ \ref{fig:wfden} (a)]. Distance $\rho \sim 15$ is 
well beyond the typical nuclear size; for configurations with such typical 
$\rho$ values the individual valence nucleons have on average 10 fm distance to 
the core.

\item Radial stabilization of the $[s_{1/2}]^2$ WF (the distances at which the 
$\sim \exp [i\varkappa \rho]$ behavior is mainly achieved) is taking place at 
quite large distances [$\rho \approx 300-400$ fm, see Fig.\ \ref{fig:wfden} 
(a)]. This is connected both to very low energy of the peak (4 keV in this 
particular calculation) and effectively long-range character of interactions in 
the $s$-wave $^8$He-$n$ channel, responsible for the formation of the three-body 
virtual state.

\item One can see from Fig.\ \ref{fig:s-inc} (c) that the cross section behavior 
in the energy region of $[p_{1/2}]^2$ state ($E \approx 2-3$ MeV) is practically 
not sensitive to the low-energy behavior of the spectrum (presence or absence of 
the $[s_{1/2}]^2$ state). It is thus possible to think that these configurations 
are practically independent in this energy region. A comparison of Figs.\ 
\ref{fig:wfden} (b) and (c) shows that this is not true. Both the internal 
structure of the WF and correlations for decay products demonstrate strong 
sensitivity to the presence of $[s_{1/2}]^2$ state, although it is not clearly 
seen in the total production cross section in this energy range.

\end{enumerate}

\begin{figure*}
\includegraphics[width=0.96\textwidth]{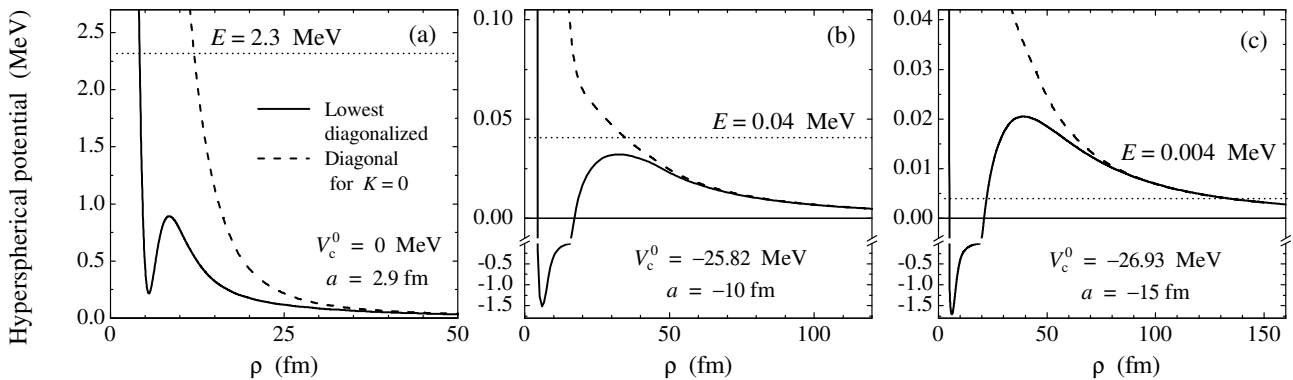} 
\caption{Hyperspherical potentials (\ref{hhpot}) as function of hyperradius. 
Solid line shows the lowest diagonalized potential, dashed line shows the lowest 
diagonal potential (in all the cases this is $K=0$ term). (a) The 
$[p_{1/2}]^2$ state at $E=2.3$ MeV [see solid curves in Fig.\ \ref{fig:s-inc} 
(c), (f)]. Panels (b) and (c) correspond to $[s_{1/2}]^2$ states at $E=0.04$ MeV 
[see short-dashed curves in Fig.\ \ref{fig:s-inc} (c), (f)] and at $E=0.004$ MeV 
respectively.}
\label{fig:poten}
\end{figure*}

An important technical insight could be obtained from the behavior of the 
effective hyperspherical interactions  (\ref{hhpot}). In Fig.\ \ref{fig:poten} 
we show the ``most attractive'' diagonal potential (this appear to to be $K=0$ 
term in all the cases) and the lowest diagonalized potential. The later can be 
considered as an effective interaction for some simple adiabatic approximation 
to the problem, which is qualitatively illustrative, but quantitatively could be 
not very reliable. The important dynamical aspects which become clear from these 
plots have already been discussed in our work \cite{gri03b} on example of broad 
states of $^5$H system. It can be seen in  Fig.\ \ref{fig:poten} that even the 
``most attractive'' diagonal potentials are in reality repulsive and do not even 
show any sign of ``pocket'' formation. The structures in the 
continuum are formed here only by interaction of multiple channels. It is 
interesting to note that the possibility of this class of states has been 
discussed many years ago (so called ``resonances of the second kind'' 
\cite{baz76}) but now we seem to face them systematically in the few-body 
dripline nuclei.

In the broad range of the resonance energies the state is located above the 
effective barrier top. Formation of the peaks, which could be very narrow (see 
Figs.\ \ref{fig:s-inc}, \ref{fig:gam-ot-e}) is presumably connected here not 
with barrier penetration, but with slow motion above the barrier and reflection 
from the right slope of the barrier \footnote{The attractive potentials with 
barriers are not necessarily needed to form resonances in the continuum. They 
can be formed by pure repulsive potentials with broad ``shelves''.}. Only in the 
extreme low energy case Fig.\ \ref{fig:poten} (c) the process could be 
interpreted as penetration through the effective barrier. Typical range of the 
barrier is consistent in this case with the ``channel radius'' estimates by Eq.\ 
(\ref{gamm-3}).

The evolution of the nuclear structure near the threshold is also an interesting 
question which is briefly discussed below. The short-dash curves ($V^0_c=-25.82$ 
MeV, $a=-10$ fm) in Fig.\ \ref{fig:s-inc} (a) and (d) show qualitatively 
different behavior compared to the expected sharpening of the threshold peak. It 
happens because in this case the bound $0^+$ state of $^{10}$He is formed with 
binding energy $\approx 60$ keV. Therefore these curves do not represent a valid 
result in the context of our studies (we should consider only the three-body 
Hamiltonians which do not lead to the bound $^{10}$He). However, it is 
interesting to see how the nuclear structure evolves in this case (Table 
\ref{tab:struc}). The virtual three-body state shows strong domination of the 
$[s_{1/2}]^2$ component in the internal region (the first row in Table 
\ref{tab:struc}). As soon as the state becomes bound, the structure changes 
drastically with rapid increase of the $[p_{1/2}]^2$ configuration weight (see 
the second row from Table \ref{tab:struc}). If we bind the $^{10}$He even more 
stronger, so that the binding energy becomes $0.3$ MeV, which corresponds to the 
binding energy of $^{11}$Li, its structure begins to resemble closely the 
typical structure of $^{11}$Li with practically equal population of the 
$[s_{1/2}]^2$ and $[p_{1/2}]^2$ configurations (the third row in Table 
\ref{tab:struc}). So, the bound analogue of the virtual three-body state is 
expected to be not a state with dominant $[s_{1/2}]^2$ configuration, but a 
state with strong ``competition'' between the $s$-wave and $p$-wave 
configurations. Usually the structure of narrow resonant (or quasibound) states 
is characterized by a high identity with the structure of the corresponding 
bound states \footnote{Good example is the structure of the  bound and 
quasibound states --- isobaric partners. An exception here is the situation of 
the Thomas-Ehrmann shift, when significant deviations from isobaric symmetry can 
be observed.}. The virtual three-body state demonstrates the behavior, which is 
qualitatively different in this respect.

Based on the presented results we can probably conclude that in the sense of 
nuclear dynamics the three-body virtual states are something intermediate 
between two-body virtual state and an ordinary resonance.

\begin{table}[b]
\caption{Internal structure of the virtual three-body state and the
bound $^{10}$He states close to the $^8$He+$n$+$n$ threshold.}
\begin{ruledtabular}
\begin{tabular}[c]{ccccc}
 $E$ (MeV) & $[s_{1/2}]^2$  & $[p_{1/2}]^2$ & $[p_{3/2}]^2$ & $[d]^2$ \\
\hline
 $0.04$\footnotemark[1] & 93.3 & 2.2 & 1.8 & 1.8 \\
 $-0.06$\footnotemark[2] & 66.1 & 23.8 & 4.7 & 4.2 \\
 $-0.3$\footnotemark[3] & 51.0 & 35.9 & 6.1 & 5.7 \\
\end{tabular}
\end{ruledtabular}
\label{tab:struc} \footnotetext[1]{Calculation of Fig.\ \ref{fig:s-inc} (c), 
short-dashed curve ($V^0_c=-25.82$ MeV, $a=-10$ fm, $V^1_c=-4.5$ MeV). Radius 
for internal normalization was taken $\rho_{\text{int}}=40$ fm, see Fig.\ 
\ref{fig:wfden} (a).} \footnotetext[2]{Continuum above this bound state is shown 
in Figs.\ \ref{fig:s-inc} (a) and (d) by the short-dashed curves ($V^0_c=-25.82$ 
MeV, $a=-10$ fm, $V^1_c=-10$ MeV).} \footnotetext[3]{The same calculation as 
\footnotemark[2], but with extra binding added by attractive three-body 
potential.}
\end{table}


\subsection{Consistence with $3 \rightarrow 3 $ scattering calculations}


The model cross section calculations for realistic $^9$He energies Fig.\
\ref{fig:s-inc} (c) and (f) show very diverse results in the case of narrow and
broad sources. A question can be asked in that case what should be considered as
a ``real'' position of $^{10}$He g.s.\ and whether it is reasonable at all to
speak about such ``real'' position if diverse experimental responses could be
expected. The theoretical approach which is ``neutral'' with respect to possible
reaction mechanism is represented by $3 \rightarrow 3 $ scattering calculations.

\begin{figure}
\includegraphics[width=0.42\textwidth]{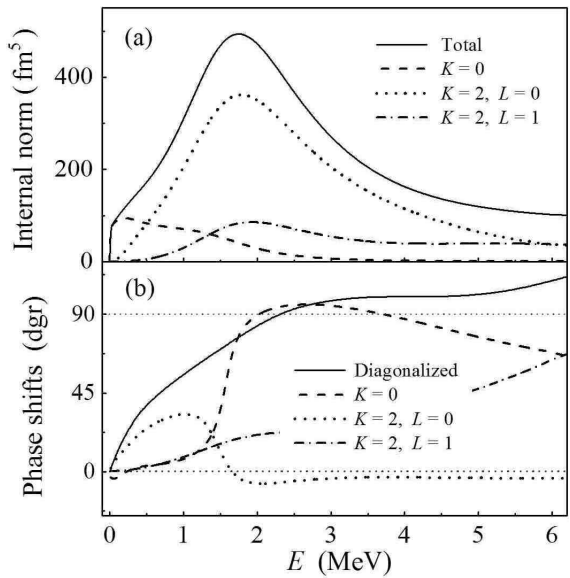}
\caption{The $3 \rightarrow 3$ scattering calculations, $V_c^0=0$, $V_c^1=-4.5$
MeV. (a) Internal normalizations Eq. (\ref{int-norm}) for dominant components of 
the WF. (b) Diagonalized phase shift (``eigenphase'') is shown by solid curve, 
while diagonal phase shifts for the lowest hyperspherical components are given 
by dashed, dotted, and dash-dotted curves.}
\label{fig:scat33-1}
\end{figure}

\begin{figure}
\includegraphics[width=0.425\textwidth]{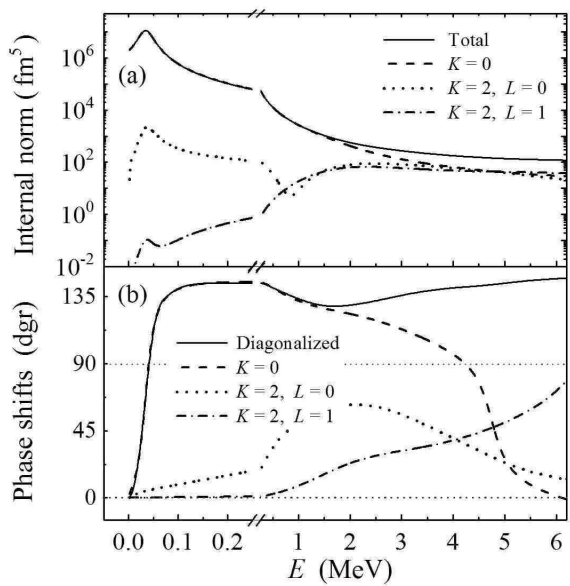}
\caption{The $3 \rightarrow 3$ scattering calculations, $V_c^0=-25.82$ MeV
($a=-10$ fm), $V_c^1=-4.5$ MeV. See Fig.\ \ref{fig:scat33-1} for details.}
\label{fig:scat33-2}
\end{figure}

Figures \ref{fig:scat33-1} and  \ref{fig:scat33-2} show the results of the $3 
\rightarrow 3$ scattering calculations in the cases of a pure $[p_{1/2}]^2$ 
state and the same in the presence of the low-energy $[s_{1/2}]^2$ state 
respectively. The details of the approach can be found in Ref.\ \cite{shu00} on 
example of the $^5$H nucleus. Three-body Hamiltonian here is the same as in 
Figs.\ \ref{fig:s-inc} (c) and (f). Three different values are displayed for 
these calculations: the diagonal $3 \rightarrow 3$ scattering phase shifts, the 
first diagonalized phase shift (so called eigenphase), and the diagonal internal 
normalizations for scattering WFs
\begin{equation}
N_{\rho_{\text{int}}}(E) = \frac{1}{\varkappa^5}\sum_{K \gamma}
\int_0^{\rho_{\text{int}}} d \rho \left| \chi_{K \gamma}^{K \gamma} (\varkappa
\rho) \right|^2 \;.
\label{int-norm}
\end{equation}
The size of the ``internal region'' $\rho_{\text{int}}=6$ fm was taken for this 
value as in Ref.\ \cite{shu00}.

For $[p_{1/2}]^2$ state the $3 \rightarrow 3$ calculations in Fig.\ 
\ref{fig:scat33-1} give somewhat different resonant energies for different 
responses: $E=1.8$ MeV for internal normalization, $E \sim 2$ MeV for the most 
strongly changing diagonal phase shifts and $E=2.3$ for eigenphase. Such spread 
is clearly connected to the fact that the phase shifts barely pass 90 degrees. 
So, we can speak about resonant energy of about $E \sim 2.0-2.3$ MeV, when only 
scattering is concerned. The agreement of $3 \rightarrow 3$ calculations with 
narrow source calculations is reasonable. That could be an indication that 
``ordinary'' reactions (simulated in this model) are a preferable tool to access 
properties of $^{10}$He compared to reactions with exotic nuclei (like 
$^{11}$Li).

In the case of narrow low-lying $[s_{1/2}]^2$ state (shown in Fig.\ 
\ref{fig:scat33-2}) the results provided by all $3 \rightarrow 3$ calculations 
(resonance energy $E=40$ keV) are in excellent agreement with each other and 
with previous model calculations [Fig.\ \ref{fig:s-inc} (c) and (f), curves with 
$a=-10$ fm]. This state is formed exclusively by $K=0$ WF component. An evidence 
for the $[p_{1/2}]^2$ state contributions could be found in the phase shift at 
around 2 MeV, but it is not very expressed. Better evidence is provided by 
internal normalizations for $K=2$ components of the WF. These show maximum at 
about 2.3 MeV and provide much broader structures, than in the case of the 
$[p_{1/2}]^2$ state not affected by $[s_{1/2}]^2$ configuration (Fig.\ 
\ref{fig:scat33-1}). Again, we can come to a conclusion that the  $[p_{1/2}]^2$ 
state survives in the presence of the  $[s_{1/2}]^2$ state in somewhat modified 
(shifted up and broadened) form, but the population of it is expected to be 
poor.

\section{Discussion}



\subsection{What is the ground state of $^{10}$He?}
\label{sec:exp-consist}


It was proposed in Ref.\ \cite{aoy02} that the observed so far state of 
$^{10}$He is not the ground but the first excited state with $[p_{1/2}]^2$ 
structure while the ground $[s_{1/2}]^2$ state remains unobserved. We confirm 
here the finding of Ref.\  \cite{aoy02} that for considerable $s$-wave 
attraction in $^{9}$He subsystem two $0^+$ states with different structures 
should coexist in the low-energy spectrum of $^{10}$He. However, we also find 
that population of the $[s_{1/2}]^2$ configuration (in the case of a strong 
$s$-wave attraction in $^{9}$He and realistic reaction scenario) is always very 
pronounced compared to the $[p_{1/2}]^2$ configuration. For that reason we can 
expect that if the $[s_{1/2}]^2$ state \emph{really exists} then the 
$[p_{1/2}]^2$ component is difficult to observe in experiment (as it is lost on 
a ``nonresonant background'' of $[s_{1/2}]^2$ low-energy excitation). It can be 
found that the energy position of the $[p_{1/2}]^2$ component of $0^+$ state is 
quite stable when the $s$-wave attraction is increased. However, for extreme 
cases of the $s$-wave attraction this contribution becomes much broader and in 
general ``lost'' on a thick right ``tail'' of the $[s_{1/2}]^2$ ground state.

\begin{figure}
\includegraphics[width=0.42 \textwidth]{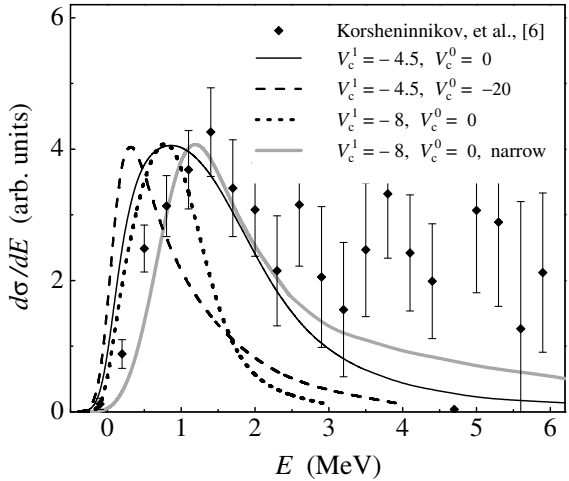}
\caption{Calculation results convoluted with experimental resolution of Ref.\
\cite{kor94} and experimental data. Solid, dashed, and dotted curves
correspond to calculations with $^{11}$Li source [see Fig.\ \ref{fig:s-inc} (f)
solid, (f) dotted, and (e) solid curves]. Gray line shows calculation with
narrow source [Fig.\ \ref{fig:s-inc} (b) dotted curve].}
\label{fig:exp-com}
\end{figure}

Current experimental situation in $^{10}$He is clearly not in favour of the 
$[s_{1/2}]^2$ state existence. Several theoretical spectra of $^{10}$He are 
provided in Fig.\ \ref{fig:exp-com} on top of the experimental data Ref.\ 
\cite{kor94}. Theoretical curves are convoluted with energy resolution of the 
experiment \cite{kor94} which is parameterized as $\Delta E= 0.7\sqrt{E}$ 
($\Delta E$ is FWHM). The calculation with $^9$He subsystem having $p_{1/2}$ 
g.s.\ at about 2 MeV reasonably fit the data. The experimental cross section 
peaked at about 1.2 MeV could be consistent with some range of $p$-wave 
interactions for $^{11}$Li source [Fig.\ \ref{fig:s-inc}, (e)--(f)]. This, 
however, is possible only for quite a weak attractive part of $s$-wave 
potential: $V^0_c > - 20$ MeV. For such value of parameters the $s$-wave 
interaction is in general still effectively repulsive (due to a large repulsive 
core). For that reason, if we completely rely on the data \cite{kor94} we would 
impose theoretical limit $a > -5$ fm for $^8$He-$n$  scattering length.  The 
derived theoretical limit is in a strong contradiction with the \emph{upper} 
limit for scattering length in $^9$He $a<-10$ fm imposed in experiment 
\cite{che01}. There is no contradiction between our theoretical limit and data 
\cite{gol07} where a \emph{lower} limit $a>-20$ fm for scattering length is 
given.

The unclear situation with exotic reaction mechanism expected for reactions with 
$^{11}$Li could have been resolved by a different experimental approach. Such an 
experiment in principle exists: the ground state of $^{10}$He and two excited 
states were identified in the complicated $2p$-$2n$ exchange reaction 
$^{10}$Be($^{14}$C,$^{14}$O)$^{10}$He \cite{ost94,boh99}. Unfortunately, the 
observed peaks rest on a ``thick'' background and has a low statistical 
confidence. None of our calculations are consistent with the results of this 
experiment. Namely, we can not reproduce in any model assumptions the small 
width of the g.s.\ obtained in this work ($300 \pm 200$ keV at 1.07 MeV of 
excitation). For example, in Fig.\ \ref{fig:s-inc} (b) the width of the state 
found at about 1.1 MeV is $\Gamma \sim 1.1$ MeV (see also Fig.\ 
\ref{fig:gam-ot-e}). Smaller width of the $^{10}$He g.s.\ \emph{if it takes 
place in reality} should mean non single-particle nature of this state [means 
not described as $^8$He(g.s.)+$n$+$n$] and hence a limited applicability of our 
model.


\subsection{Prospects of correlation studies}
\label{sec:corel}


Important structure information about the three-body system could be obtained
analysing the correlations among the decay products. The recent examples of such
data analysis include successful application to the broad states in the
continuum of $^5$H system \cite{gol05b} and to the two-proton radioactivity
decays of $^{19}$Mg \cite{muk07} and $^{45}$Fe \cite{mie07}. Such range of
application indicates a potential power of the correlation studies. 

The partial decompositions of the cross section given in Fig.\ \ref{fig:bn-com} 
show how the contributions of the $[s_{1/2}]^2$ component (mainly $K=0$) and 
$[p_{1/2}]^2$ component (mainly $K=2$) change when we add the $s$-wave 
interaction in $^9$He channel on top of the $p$-wave interaction or switch from 
the narrow to the broad source. The qualitative differences in these 
decompositions should be seen as qualitative differences in the correlation 
patterns.

\begin{figure*}
\includegraphics[width=0.47 \textwidth]{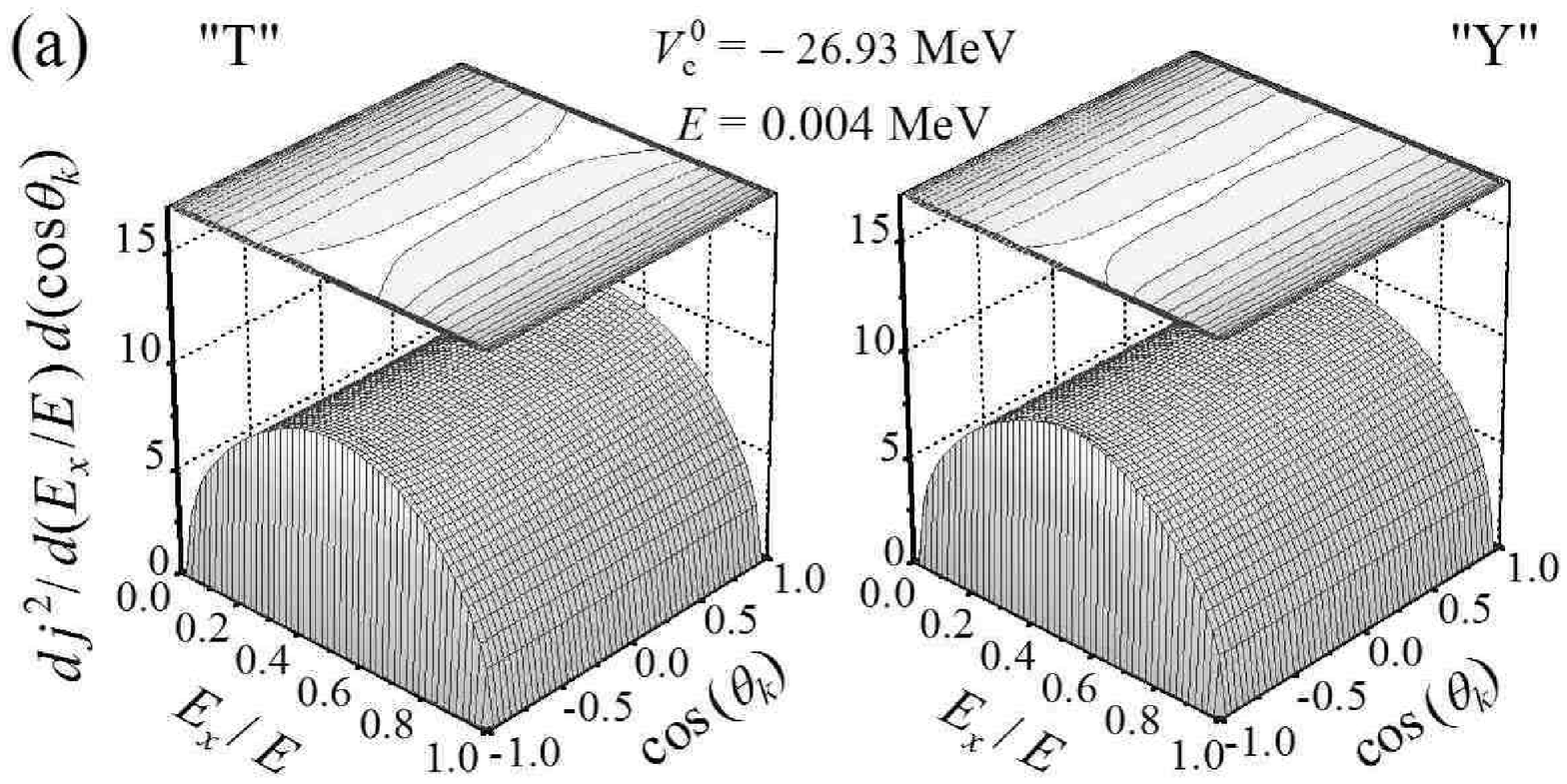}~
\includegraphics[width=0.47 \textwidth]{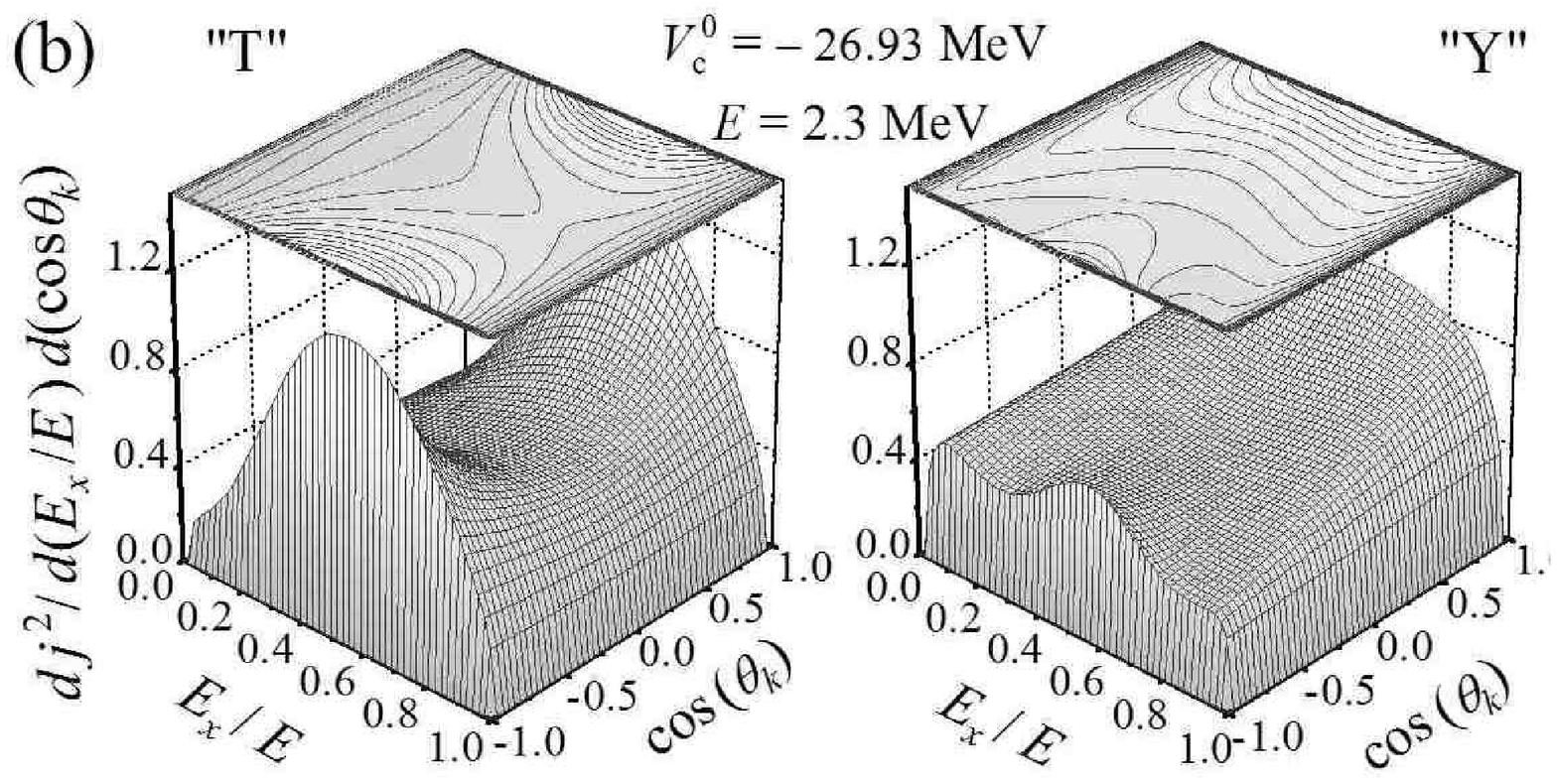} \\
\includegraphics[width=0.47 \textwidth]{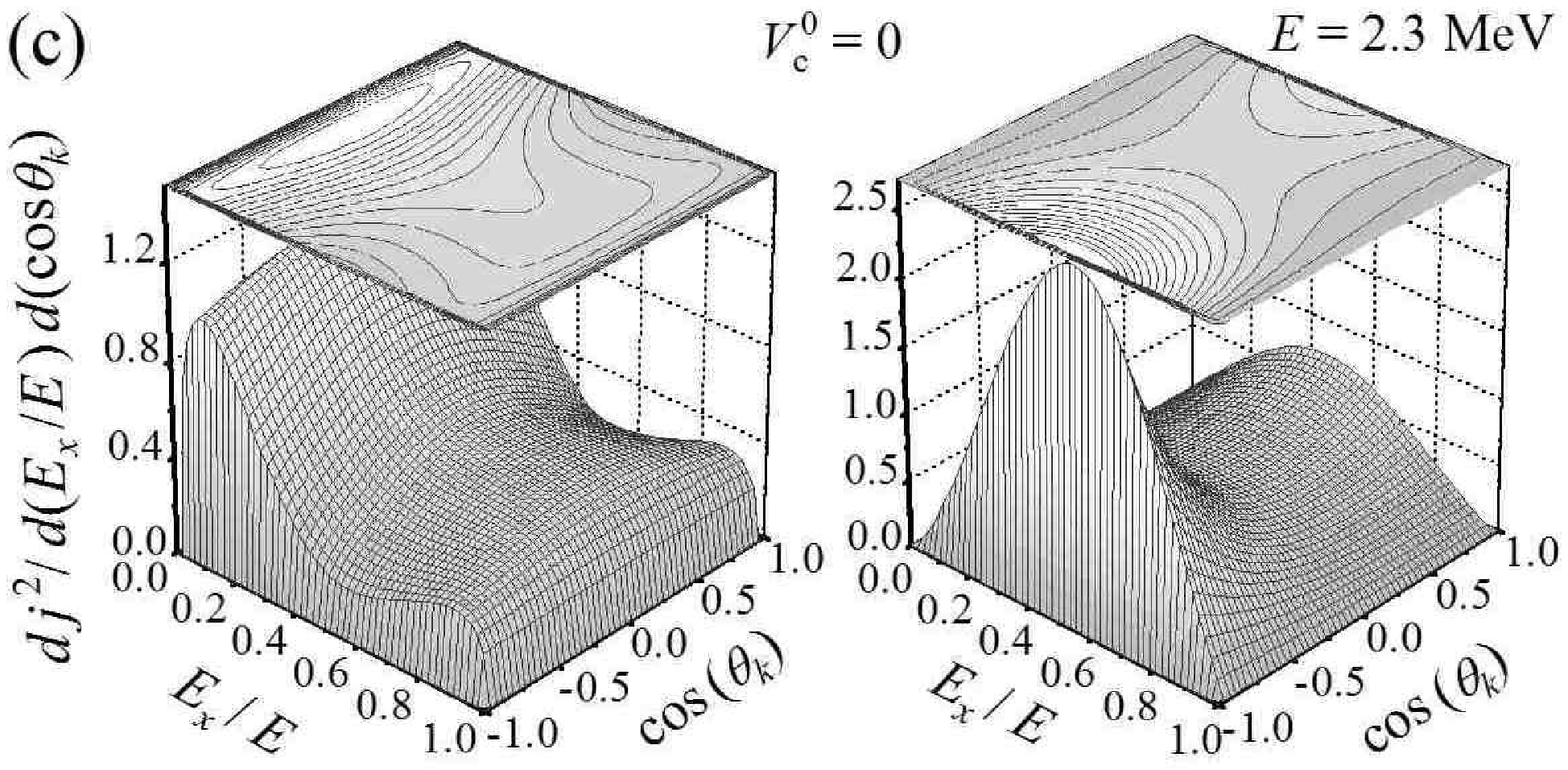}~
\includegraphics[width=0.47 \textwidth]{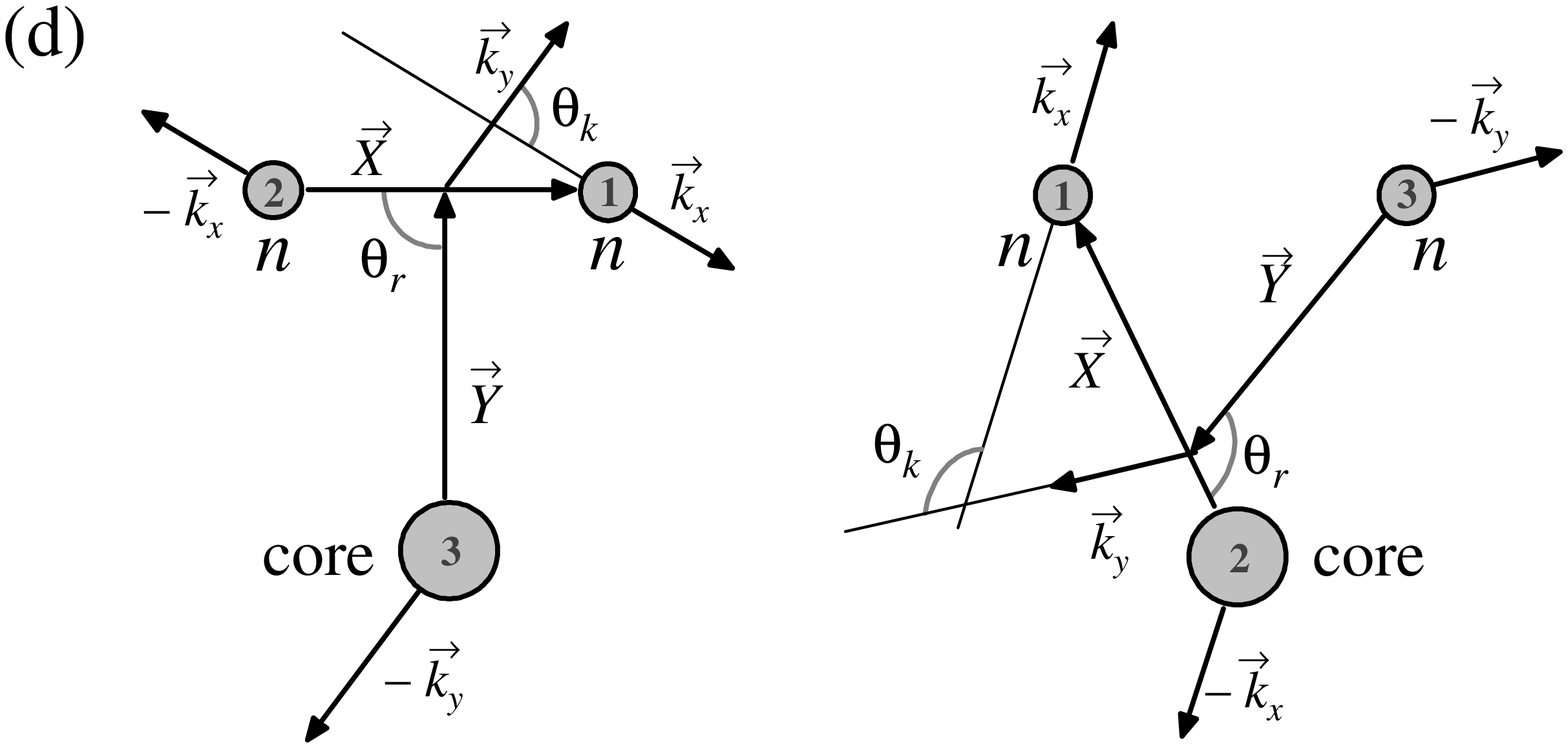}
\caption{Complete correlation information about decays of $0^+$ state at given
energy $E$. Left and right columns show the same result in ``T'' and ``Y''
Jacobi systems. Calculations provided in panels (a), (b), and (c) correspond to 
the same panels of Fig.\ \ref{fig:wfden}. Panel (d) shows the coordinates and 
angles in radial and momentum space for the ``T'' and ``Y'' Jacobi systems.}
\label{fig:corel}
\end{figure*}

The complete correlation information is provided in Fig.\ \ref{fig:corel} for
correlations in the $[s_{1/2}]^2$ state [$E=4$ keV, panel (a)] and in the
$[p_{1/2}]^2$ state at $E=2.3$ MeV in the calculations with attraction in
$s$-wave (b) and no $s$-wave attraction (c). The correlation densities are given
in the plane of parameters $\{ \varepsilon, \cos(\theta_k) \}$ both in ``T'' and 
``Y'' Jacobi coordinate systems. Parameter $\varepsilon=E_x/E$ describes the 
energy distribution between $X$ and $Y$ Jacobi subsystems ($E_x$ is energy in 
the $X$ Jacobi subsystem). Parameter $\theta_k$ is angle between Jacobi momenta
$k_x$ and $k_y$ in the chosen Jacobi coordinate system:
\[
\cos(\theta_k) = \frac{(\mathbf{k}_x,\mathbf{k}_y)}{ k_{x}\;  k_y}\;.
\]
More details can be found in Ref.\ \cite{gri03c}.

The correlation picture for the virtual three-body state with the $[s_{1/2}]^2$
structure is quite featureless [Fig.\ \ref{fig:corel} (a)]. The energy
distribution between subsystems is close to the ``phase volume'' distribution
\begin{equation}
\frac{d^2 \sigma}{ dE d E_{x} } \sim E \;\sqrt{E_x(E-E_x)} \;.
\label{eq:ps}
\end{equation}
There are only minor deviations from flat distribution for $\cos(\theta_k)$ at
$\theta_k \sim 0^{\circ}$ and $\theta_k \sim 180^{\circ}$ in ``T'' Jacobi system
(these are configurations when three particles come out in a line).

The predicted correlations for the $[p_{1/2}]^2$ state in ``T'' Jacobi system 
[Fig.\ \ref{fig:corel} (c)] look very much like those already observed in other 
$p$-wave systems $^6$Be \cite{boc89} and  $^5$H \cite{gol05b}. There is a 
double-hump structure reflecting the $[p_{1/2}]^2$ population. The  hump, which 
corresponds to low-energy motion between neutrons  is strongly enhanced due to 
FSI in the $n$-$n$ cannel.

Fig.\ \ref{fig:corel} (b) provides prediction of correlations which is possibly 
important for prospective $^{10}$He studies. If the attractive $s$-wave 
interaction is added it qualitatively changes the picture of correlations at the 
expected  $[p_{1/2}]^2$ state position. Now the energy distribution between 
subsystems (in the  ``T'' Jacobi system) is close to the phase space 
distribution Eq.\ (\ref{eq:ps}). Also the angular distribution changes 
drastically: the correlation density is concentrated in the regions where one of 
the neutrons is close to the $^8$He core in the momentum space [$\cos(\theta_k) 
\sim \pm 1$ and $E_x/E \sim 5/9$ in the ``T'' Jacobi system]. In this case the 
only expressed feature in the ``Y'' Jacobi system is the ``dineutron'' 
correlation which can be seen as a small peak at $\cos(\theta_k) \sim - 1$ and 
$E_x/E \sim 1/2$.

The drastic changes between distributions Fig.\ \ref{fig:corel} (b) and (c) mean 
that in the experimental measurements giving access to such a characteristic 
there will be no doubts in the structure identification even in the case of a 
poor population of the low-energy part of the spectrum or technical problems 
with detection of the low-energy events.


\subsection{Reliability of the results}


It should be mentioned once again that the aspects of the $^{10}$He dynamics, 
discussed in this work, are only valid if the single-particle 
$^{8}$He(g.s.)+$n$+$n$ structure of the low-lying $^{10}$He states really takes 
place. The ground for such an assumption is provided by knowledge of the 
$^{9}$He spectrum. However, the narrow first resonant states of $^{9}$He, as 
observed in Refs.\ \cite{set87,boh88} [$E(p_{1/2})=1.27$, MeV, $\Gamma=0.1$ MeV 
and $E(p_{3/2})=2.4$ MeV, $\Gamma=0.7$ MeV], presume that it is not true, 
because small spectroscopic factors are expected \cite{bar04}. In the case that 
the results of Ref.\ \cite{gol07} are preferable [$E(p_{1/2})=2$ MeV, $\Gamma=2$ 
MeV], implying that this is a single-particle state, the basis for our model 
becomes reliable.

Sensitivity of the predictions to the $s$-wave interaction in the $^9$He channel 
is very high. The experimental results of Refs.\ \cite{che01} ($a<-10$ fm) and 
\cite{gol07} ($a>-20$ fm) are not contradictory, although not too restrictive. 
Thus, still no solid experimental ground can be found here. We think that this 
issue is a key point for understanding of the $^{10}$He structure.

Important conclusion of these studies is that energy spectrum obtained in 
experiments with $^{11}$Li is strongly affected by the reaction mechanism and we 
do not reproduce the results of the experiment \cite{kor94} without taking this 
effect into account. The question can be raised from theoretical side, how 
reliable is the statement that for $p_{1/2}$ state in $^9$He at about 2 MeV we 
can not get a \emph{state} in $^{10}$He at 1.2 MeV straightforwardly. In Table 
\ref{tab:paring} we list paring energy $E_p$ for valence neutrons calculated for 
$^{10}$He in different theoretical approaches. With $p_{1/2}$ state at 2 MeV the 
paring energy should be about 2.8 MeV, while in various theoretical approaches 
it is typically around 1 MeV and never exceeds 2 MeV. It is clear that 
relatively small paring energy in $^{10}$He is common for different theoretical 
models and can be considered as a reliable prediction. Also if we have a look on 
the nearby isotopes, for $p_{3/2}$ subshell nuclei $^{6}$He and $^{8}$He $E_p$ 
is $2.6$ MeV and $3.04$ MeV respectively. For $^{11}$Li, where $p_{1/2}$ 
subshell is populated, paring energy is known to be small: $E_p \sim 0.8$ MeV.


\section{Conclusion}


In conclusion we would like to emphasize the most important results of our
studies:

\noindent (i) Within theoretical model for $p_{1/2}$ state in the $^9$He located
at about 2 MeV it is problematic to obtain the $^{10}$He g.s.\ at about 1.2 MeV
straightforwardly. The required for that paring energy is 2.8 MeV, while for
$[p_{1/2}]^2$ configuration it is typically obtained $\sim 1-2$ MeV.

\noindent (ii) The attraction in $s$-wave allows to shift state with 
$[p_{1/2}]^2$ configuration to significantly lower energy. However, for some 
extreme values of attraction ($a \leq -5$ fm) lead to formation of low-energy 
$[s_{1/2}]^2$ state which is seen as a sharp peak in the cross section at 
energies less than 0.3 MeV. The appearance of such a state is in accord with 
predictions of Ref.\ \cite{aoy02}.

\noindent (iii) In contrast with approach of Ref.\ \cite{aoy02}, we study the
conditions of ``coexistence'' of $[s_{1/2}]^2$ and $[p_{1/2}]^2$ states in the
$0^+$ continuum for realistic scenarios. It is shown that the state with
$[p_{1/2}]^2$ structure is poorly populated (also suffer significant broadening)
in the presence of $[s_{1/2}]^2$ ground state and can be easily lost (small on
the $s$-wave ``background''). For that reason the idea of Ref.\ \cite{aoy02}
that the ground $[s_{1/2}]^2$ state of the $^{10}$He remains unobserved, while
the observed so far state is the first excited one with $[p_{1/2}]^2$ structure,
does not get support in our studies.

Concerning comparison with experimental data:

\noindent (i) Observation of quite a broad peak in $^{10}$He at about 1.2 MeV in 
Ref.\ \cite{kor94} could be explained  by a specific mechanism of the chosen 
reaction induced by $^{11}$Li (namely the huge size of neutron halo in 
$^{11}$Li). This explanation is possible only in the case of absence of virtual 
state in $^9$He channel. For $^{10}$He ground $[p_{1/2}]^2$ state located at $E 
\geq 2$ MeV the mentioned reaction mechanism leads to a strong enhancement of 
the low-energy transition strength even without any significant attraction in 
$s$-wave. As a result, the peak in the cross section may be shifted to a lower 
energy (e.g.\ $\sim 1.2$ MeV).

\noindent (ii) The provided theoretical model essentially infer the properties
of the $^{10}$He system basing on the properties of the $^9$He subsystem. At the
moment we can not make the existing data on $^9$He and $^{10}$He consistent
within this model. Calculations with large negative scattering length (e.g.\
$a<-5$ fm; experimental limit \cite{che01} is $a<-10$ fm) in core-$n$ subsystem
necessarily lead to formation of the single narrow peak below 0.3 MeV in the
spectrum which should have been seen in the experiment \cite{kor94}.

\noindent (iii) We have to conclude that the existing experimental data do not
allow to establish unambiguously the ``real'' g.s.\ position for $^{10}$He.
Alternative experiments (relative to those utilizing $^{11}$Li beams) are
desirable. Further clarification of controversy between the $^9$He and $^{10}$He
spectra is indispensable for theoretical understanding of the Helium isobar
properties.

%
\section{Acknowledgements} 
%
%

We are grateful to Prof.\ Yu.\ Ts.\ Oganessian for inspiration for this work. We 
are grateful for careful reading of the manuscript and valuable discussions to 
Profs.\ A.\ A.\ Korsheninnikov, G.\ M.\ Ter-Akopian, and M.\ S.\ Golovkov. LVG 
is supported by the INTAS Grant 05-1000008-8272, Russian RFBR Grants Nos.\ 
05-02-16404 and 05-02-17535, and Russian Ministry of Industry and Science grant 
NS-1885.2003.2.



\begin{table*}[b]
\caption{Paring energy (in MeV) for $^{10}$He defined as $E_p=S_{2n}-2S_n$
calculated in different theoretical approaches.}
\begin{ruledtabular}
\begin{tabular}[c]{cccccccc}
Work & \cite{ste88} & \cite{kor93}\footnotemark[1]  & \cite{she96} &
\cite{nav98} & \cite{aoy02} & \cite{vol05}\footnotemark[2]   &  This
work\footnotemark[1]\\
\hline
$-S_{n}$  & 1.22 & 0.74 & 0.84 & 2.38 & 1.27 & 1.60 & 2.0  \\
$-S_{2n}$ & 1.18 & 0.6  & 1.09 & 2.78 & 1.68 & 1.94 & $\sim 2.0-2.3$  \\
$E_p$     & 1.26 & 0.88 & 0.59 & 1.98 & 0.86 & 1.25 & $\sim 1.7-2.0$  \\
\end{tabular}
\end{ruledtabular}
\label{tab:paring}
\footnotetext[1]{We use $p_{1/2}$ elastic cross section peak energy to define
$S_n$.}
\footnotetext[2]{See Table 1, column 6 of Ref.\ \cite{vol05}.}
\end{table*}


\begin{thebibliography}{99}



\bibitem{baz69}
A.\ I.\ Baz, V.\ F.\ Demin, and M.\ V.\ Zhukov,
Sov.\ J.\ Nucl.\ Phys.\ \textbf{9}, 693 (1969) [Yad.\ Fiz.\ \textbf{9}, 1184
(1969)].


\bibitem{ste88}
J.\ Stevenson, B.\ A.\ Brown, Y.\ Chen, J.\ Clayton, E.\ Kashy, D.\ Mikolas, J.\
Nolen, M.\ Samuel, B.\ Sherrill, J.\ S.\ Winfield, Z.\ Q.\ Xie, R.\ E.\ Julies,
W.\ A.\ Richter,
Phys.\ Rev.\ C \textbf{37}, 2220 (1988).


\bibitem{set87}
K.\ K.\ Seth, M.\ Artuso, D.\ Barlow, S.\ Iversen, M.\ Kaletka, H.\ Nann, B.\
Parker, R.\ Soundranayagam,
Phys.\ Rev.\ Lett.\  \textbf{58}, 1930 (1987).


\bibitem{boh88}
H.\ G.\ Bohlen, B.\ Gebauer, D.\ Kolbert, W.\ von Oertzen, E.\ Stiliaris, M.\
Wilpert, T.\ Wilpert,
Z.\ Phys.\ \textbf{A330}, 227 (1988).


\bibitem{kor93}
A.\ A.\ Korsheninnikov, B.\ V.\ Danilin, and M.\ V.\ Zhukov,
Nucl.\ Phys.\ \textbf{A559}, 208 (1993).


\bibitem{kor94}
A.\ A.\ Korsheninnikov, K.\ Yoshida, D.\ V.\ Aleksandrov, N.\ Aoi, Y.\ Doki, N.\
Inabe, M.\ Fujimaki, T.\ Kobayashi, H.\ Kumagai, C.-B.\ Moon, E.\ Yu.\
Nikolskii, M.\ M.\ Obuti, A.\ A.\ Ogloblin, A.\ Ozawa, S.\ Shimoura, T.\ Suzuki,
I.\ Tanihata, Y.\ Watanabe, M.\ Yanokura,
Phys.\ Lett.\ \textbf{B326}, 31 (1994).


\bibitem{ost94}
A.\ N.\ Ostrowski, H.\ G.\ Bohlen, B.\ Gebauer, S.\ M.\ Grimes, R.\
Kalpakchieva, Th.\ Kirchner, T.\ N.\ Massey, W.\ von Oertzen, Th.\ Stolla, M.\
Wilpert, Th.\ Wilpert,
Phys.\ Lett.\ \textbf{B338}, 13 (1994).


\bibitem{boh99}
H.\ G.\ Bohlen, A.\ Blazevic, B.\ Gebauer, W.\ Von Oertzen, S.\ Thummerer,
R. Kalpakchieva, S.\ M.\ Grimes, and T.\ N.\ Massey,
Prog.\ Part.\ Nucl.\ Phys.\ \textbf{42}, 17 (1999).


\bibitem{kob97}
T.\ Kobayashi, K.\ Yoshida, A.\ Ozawa, I.\ Tanihata, A.\ Korsheninnikov,
E.\ Nikolski, and T.\ Nakamura,
Nucl.\ Phys.\ \textbf{A616}, 223c (1997).


\bibitem{che01}
L.\ Chen, B.\ Blank, B.\ A.\ Brown, M.\ Chartier, A.\ Galonsky, P.\ G.\ Hansen,
M.\ Thoennessen,
Phys.\ Lett.\  \textbf{B505}, 21 (2001).


\bibitem{aoy02}
S.\ Aoyama,
Phys.\ Rev.\ Lett.\ \textbf{89}, 052501 (2002).


\bibitem{gol07}
M.\ S.\ Golovkov, L.\ V.\ Grigorenko, A.\ S.\ Fomichev, A.\ V.\
Gorshkov, V.\ A.\ Gorshkov, S.\ A.\ Krupko, Yu.\ Ts.\ Oganessian, A.\ M.\ Rodin,
S.\ I.\ Sidorchuk, R.\ S.\ Slepnev, S.\ V.\ Stepantsov, G.\ M.\ Ter-Akopian, R.\
Wolski, A.\ A.\ Korsheninnikov, E.\ Yu.\ Nikolskii, V.\ A.\ Kuzmin, B.\ G.\
Novatskii, D.\ N.\ Stepanov, P.\ Roussel-Chomaz, W.\ Mittig,
Phys.\ Rev.\ C \textbf{76}, 021605(R) (2007).


\bibitem{gog70}
D.\ Gogny, P.\ Pires and R.\ de Tourreil,
Phys. Lett. \textbf{B32}, 591 (1970).


\bibitem{gri03b}
L.\ V.\ Grigorenko, N.\ K.\ Timofeyuk, and M.\ V.\  Zhukov,
Eur.\ Phys.\ J.\ \textbf{A 19}, 187 (2004).


\bibitem{asc69}
R.\ J.\ Ascuitto and N.\ K.\ Glendenning,
Phys.\ Rev.\ \textbf{181}, 1396 (1969).


\bibitem{shu06}
N.\ B.\ Shulgina,
 private communication.


\bibitem{gri07}
L.\ V.\ Grigorenko, and M.\ V.\ Zhukov,
Phys.\ Rev.\ C \textbf{76}, 014008 (2007).


\bibitem{dan07}
B.\ V.\ Danilin, J.\ S.\ Vaagen, T.\ Rogde, S.\ N.\ Ershov, I.\ J.\ Thompson, 
and M.\ V.\ Zhukov,
Phys.\ Rev.\ C \textbf{76}, 064612 (2007).


\bibitem{glo78}
W.\ Gl\"ockle,
Phys.\ Rev.\ C \textbf{18}, 564 (1978).


\bibitem{tan99}
N.\ Tanaka, Y.\ Suzuki, K.\ Varga, R.\ G.\ Lovas,
Phys.\ Rev.\ C \textbf{59}, 1391 (1999).


\bibitem{del00}
A.\ Delfino, T.\ Frederico, and L.\ Tomio, 
Few-Body Syst.\ \textbf{28}, 259 (2000).

\bibitem{fre07}
T.\ Frederico and M.\ T.\ Yamashita,
Nucl.\ Phys.\ \textbf{A790}, 116c (2007).


\bibitem{gri01}
L.\ V.\ Grigorenko, R.\ C.\ Johnson, I.\ G.\ Mukha, I.\ J.\ Thompson, and M.\ 
V.\ Zhukov,
Phys.\ Rev.\ C \textbf{64}, 054002 (2001).


\bibitem{gol04a}
M.\ S.\ Golovkov, L.\ V.\ Grigorenko, A.\ S.\ Fomichev, Yu.\ Ts.\
Oganessian, Yu.\ I.\ Orlov, A.\ M.\ Rodin, S.\ I.\ Sidorchuk, R.\ S.\ Slepnev,
S.\ V.\ Stepantsov, G.\ M.\ Ter-Akopian, R.\ Wolski,
Phys.\ Lett.\ \textbf{B588}, 163 (2004).



\bibitem{baz76} 
A.\ I.\ Baz, 
Zh.\ Eksp.\ Teor.\ Fiz.\ \textbf{70}, 397 (1976).







\bibitem{zhu93}
M.\ V.\ Zhukov, B.\ V.\ Danilin, D.\ V.\ Fedorov, J.\ M.\ Bang, I.\ J.\
Thompson, and J.\ S.\ Vaagen,
Phys.\ Rep.\ {\bf 231}, 151 (1993).


\bibitem{shu00}
N.\ B.\ Shulgina, B.\ V.\ Danilin, L.\ V.\ Grigorenko,
M.\ V.\ Zhukov, and J.\ M.\ Bang,
Phys.\ Rev.\ C \textbf{62}, 014312 (2000)


\bibitem{gol05b}
M.\ S.\ Golovkov, L.\ V.\ Grigorenko, A.\ S.\ Fomichev, S.\ A.\
Krupko,  Yu.\ Ts.\ Oganessian, A.\ M.\ Rodin, S.\ I.\ Sidorchuk, R.\ S.\
Slepnev, S.\ V.\ Stepantsov, G.\ M.\ Ter-Akopian, R.\ Wolski, M.\ G.\ Itkis, A.\ 
S.\ Denikin, A.\ A.\ Bogatchev, N.\ A.\ Kondratiev, E.\ M.\ Kozulin, A.\ A.\ 
Korsheninnikov, E.\ Yu.\ Nikolskii, P.\ Roussel-Chomaz, W.\ Mittig, R.\ Palit, 
V.\ Bouchat, V.\ Kinnard, T.\ Materna, F.\ Hanappe, O.\ Dorvaux, L.\ Stuttg\'e, 
C.\ Angulo, V.\ Lapoux, R.\ Raabe, L.\ Nalpas, A.\ A.\ Yukhimchuk, V.\ V.\ 
Perevozchikov, Yu.\ I.\ Vinogradov, S.\ K.\ Grishechkin,
S.\ V.\ Zlatoustovskiy, 
Phys.\ Rev.\ C \textbf{72}, 064612 (2005).


\bibitem{muk07}
I.\ Mukha, K.\ S\"ummerer, L.\ Acosta, M.\ A.\ G.\ Alvarez, E.\ Casarejos, A.\
Chatillon, D.\ Cortina-Gil, J.\ Espino, A.\ Fomichev, J.\ E.\ Garc³a-Ramos, H.\
Geissel, J.\ Gomez-Camacho, L.\ Grigorenko, J.\ Hofmann, O. Kiselev, A.\
Korsheninnikov, N.\ Kurz, Yu.\ Litvinov, I. Martel, C.\ Nociforo, W.\ Ott, M.\
Pfutzner, C.\ Rodr³guez-Tajes, E.\ Roeckl, M.\ Stanoiu, H.\ Weick, and P.\ J.\
Woods,
Phys.\ Rev.\ Lett.\ \textbf{99}, 182501 (2007)


\bibitem{mie07}
K.\ Miernik, W.\ Dominik, Z.\ Janas, M.\ Pf\"utzner, L.\ Grigorenko, C.\ R.\
Bingham, H.\ Czyrkowski, M.\ Cwiok, I.\ G.\ Darby, R.\ Dabrowski, T.\ Ginter,
R.\ Grzywacz, M.\ Karny, A. Korgul, W.\ Kusmierz, S.\ N.\ Liddick, M.\ Rajabali,
K.\ Rykaczewski, and A.\ Stolz,
Phys.\ Rev.\ Lett. \textbf{99}, 192501 (2007).


\bibitem{gri03c}
L.\ V.\ Grigorenko, and M.\ V.\ Zhukov,
Phys.\ Rev.\ C \textbf{68}, 054005 (2003).


\bibitem{boc89}
O.\ V.\ Bochkarev, A.\ A.\ Korsheninnikov, E.\ A.\ Kuz'min, I.\ G.\ Mukha,
L.\ V.\ Chulkov, G.\ B.\ Yan'kov,
Nucl.\ Phys.\ \textbf{A505}, 215 (1989).


\bibitem{bar04}
F.\ C.\ Barker,
Nucl.\ Phys.\ \textbf{A741}, 42 (2004).


\bibitem{she96}
Y.\ S.\ Shen and Z.\ Ren,
Phys.\ Rev.\ C \textbf{54} 1158 (1996).


\bibitem{nav98}
P.\ Navratil and B.\ R.\ Barrett,
Phys.\ Rev.\ C \textbf{57}, 3119 (1998).


\bibitem{vol05}
A.\ Volya and V.\ Zelevinsky,
Phys.\ Rev.\ Lett.\ \textbf{94}, 052501 (2005).



\end{thebibliography}
\end{document}